\documentclass[12pt]{iopart} \usepackage{iopams} \usepackage{epsfig}
\begin{document}
\title{Water in nanopores. II. Liquid-vapour phase transition near
hydrophobic surfaces}
\author{Ivan Brovchenko\dag\, Alfons Geiger\dag\
and Alla Oleinikova\ddag}

\address{\dag\ Physical Chemistry, Dortmund University, 44221
Dortmund, Germany} \address{\ddag\ Physical Chemistry, Ruhr-University
Bochum, 44780 Bochum, Germany}
\eads{\mailto{brov@heineken.chemie.uni-dortmund.de}, \mailto{alfons.geiger@udo.edu}, \mailto{alla.oleinikova@ruhr-uni-bochum.de}}
\begin{abstract}
The liquid-vapour  phase transition near a weakly attractive
surface is studied by simulations of  the  coexistence  curves of
water  in hydrophobic pores. There is a pronounced gradual density
depletion of the liquid phase near the surface without
any trend to the formation of a vapour layer  below  the bulk critical
temperature T$_{C}$.  The temperature  dependence of the  order
parameter in the  surface layer follows the power law
($\rho_{l}$ - $\rho_{v}$)    $\sim$ (1 - T/T$_{C}$)$^{\beta_{1}}$
with a value of the exponent $\beta_{1}$ close  to the critical
exponent $\beta_{1}$ =  0.82 of  the {\it ordinary} transition  in the
Ising  model. The order parameter profiles in the subcritical region
are consistent  with the behaviour of an ordinary transition  and
their temperature evolution is governed by  the   bulk
correlation length. Density  profiles of water at  supercritical
temperatures are consistent with the behaviour of the {\it normal} transition
caused by the preferential adsorption of voids. The relation
between normal and ordinary transitions in the Ising model and in fluids is
discussed.
\end{abstract}

\submitto{\JPCM}


\section{Introduction}
Studies of the water behaviour  near hydrophobic surfaces are
necessary for the understanding of various phenomena: hydrophobic
attraction between extended surfaces, slipping flow of liquid
water near a hydrophobic surface, hydrophobic  interaction  between
large solutes in aqueous solutions, etc. In particular, the hydrophobic
interaction   plays  an  important  role  in the processes  of
protein folding. The ultimate  prerequisite of  such studies is
the knowledge of the phase behaviour of water near the surface, that
includes surface critical behaviour and location of  the surface
phase transitions.
\par In  the case of a weak fluid-wall interaction a drying
transition, i.e the formation of  a macroscopically thick vapour
layer near the surface  in the liquid phase,  could be expected
in general, similarly to  the  wetting  transition  in the  case
of  the a strong fluid-wall  interaction
\cite{Cahn,NF82,Pandit,Dietrichrev}. For a fluid near a hard wall
a drying transition, which is accompanied  by the formation of a
liquid-vapour interface at some distance from the substrate, is
expected \cite{Henderson85}. The possible  appearance  of a
vapour layer around hard spheres in liquid water was also
discussed \cite{Still,Chandler99}. In simulations a drying-like
surface phase transition in the liquid phase of water near a
substrate with a repulsive step in the interaction potential was
found \cite{BrovNATO}.
 For a weakly attractive {\it short-range} fluid-wall
interaction a drying transition, as expected from general theoretical
arguments \cite{NF82,Pandit}, was reported for Lennard-Jones (LJ)
fluids \cite{Henderson,Nijm91}.  In the case  of a {\it long-range} fluid-wall
interaction  (for  example,  via  van  der Waals  forces) a drying
transition takes place at the bulk  critical   temperature  only
\cite{EbnerSaam, Indekeu85, Chan}. This  means, that a macroscopic vapour
layer between the liquid and the surface probably never occurs in real  systems.
Indeed, a drying transition was never observed experimentally.
The absence of a drying transition also eliminates the possibility to
observe predrying transitions
in the  stable liquid states, which means the formation of a thin vapour
layer in states outside the liquid-vapour coexistence region.
 \par  The presence of a solid boundary  effects the  critical  behaviour of
a fluid. This in turn determines the properties of the fluid not only near
  the critical
point but in a wider range of the thermodynamic conditions. The fluid
becomes  heterogeneous and  the local  properties,  including the density,
depend  on  the  distance  to  the surface.  The behaviour  of  the  local
properties near the surface should follow the laws of some  surface
universality class. The
 bulk fluid belongs to the universality class of
the 3D Ising model, which shows three surface universality  classes,
depending  on  the interaction with the surface and the
interparticle interaction in the surface layer \cite{Binderrev1}.   In
fluids, the average intermolecular interaction energy per particle
always diminishes near  the surface due to the effect of missing neighbours
and,  therefore,  only  two  universality  classes  may  be  relevant:
the so-called ordinary and normal transitions.
\par In a zero surface
field h$_{1}$ =  0 (free boundary), the effect of missing neighbours results
in the {\it ordinary} transition universality class    behaviour
\cite{BinderHoh,BinderLan}.  The order parameter (magnetization
$\left|m\right|$  in the case of the Ising model)  in both  coexisting phases
significantly decreases towards the  surface at temperatures T
below the critical temperature T$_{C}$ and its profile becomes flat (m
= 0) at T$_{C}$ and in the supercritical region.
The temperature dependence of the order parameter in
the surface   layer m$_{1}$  follows   the    scaling   law \cite{BinderHoh}
\begin{eqnarray}
  m_{1} = \pm B_1  \tau^{\beta_1},
\end{eqnarray}
 where  $\tau$ =  (T$_{C}$ -  T)/T$_{C}$ is  the  reduced temperature,
 $\beta_1$  $\approx$  0.82   \cite{beta1}  is  the  surface  critical
 exponent, which is essentially higher than the bulk critical exponent
 $\beta$  $\approx$ 0.326  \cite{beta}.
\par  A non-zero  surface field
 $\left|h_{1}\right|$ $>$  0 causes preferential adsorption  of one of
 the  components and a behaviour according to the {\it normal}
 transition  universality class.   In such
 a case the  magnetization profile does  not vanish at T  $\geq$ T$_{C}$
 and the deviation of the magnetization from the bulk  value increases toward
 the  surface. In fluids this effect causes  the  phenomenon of
 critical adsorption \cite{Critads}.  The surface critical behaviour in
 the limit $\left|h_1\right|$  $\rightarrow$ $\infty$ is equivalent to
 the  so-called {\it extraordinary} transition,  which  occurs,  when  the
 interaction between the spins in the surface layer exceeds some critical
 value  \cite{Diehl}.  Although  in the case of a normal  transition the
 magnetizations of  the  coexisting   phases  are   not  symmetrical
 relative to the axis m  = 0,  the  difference  of  magnetizations
 approaches a flat profile $\Delta$m (z) = 0, when T  $\rightarrow$ T$_{C}$,
 similarly   to  the   ordinary  transition   (and  contrary   to  the
 extraordinary  transition  \cite{Binderrev1}).  For  the  normal  and
 extraordinary transitions  the magnetization in the  surface layer is
 predicted to show the following temperature dependence \cite{Diehl}:
\begin{eqnarray}
m_1 = (m_{1C} + A_1 \tau + A_2\tau^2 + ...) + B_\pm \tau^{2-\alpha},
\end{eqnarray}
 where the contribution in  parentheses is a regular background, while
 the  leading   singular  contribution  is   $\sim$  $\tau^{2-\alpha}$
 ($\alpha$ $\approx$ 0.11  \cite{beta}), when approaching T$_{C}$ from
 below (B$_-$) or above (B$_+$). Note,  that the singular  contribution  becomes
 negligible  compared to the linear regular  term when  T $\rightarrow$
 T$_{C}$.
\par In the general  case $\left|h_1\right|$  $<$ $\infty$  the
 magnetization profile at T = T$_{C}$ shows a maximum at some distance
 from the surface.  This  maximum increases and approaches the surface
 when $\left|h_1\right|$  $\rightarrow$ $\infty$, while  it moves away
 from   the   surface    and   disappears,   when   $\left|h_1\right|$
 $\rightarrow$ 0  \cite{Ritschel,Maciolek}.  Such behaviour corresponds
 to the crossover of the critical behaviour from  ordinary transition near
 the surface to normal transition  away from it, where  the crossover
 distance approximately coincides with the maximum of the magnetization
 at T = T$_{C}$.  This means that in a surface layer the critical behaviour
 corresponding to the normal transition could be observed in the limit
 of strong  adsorption only.
\par  As far as  a fluid belongs  to the
 same  universality class as the 3D Ising  model it  is natural to map
 the surface  critical behaviour of fluids to the  surface  universality
 classes  described above. The symmetry of liquid binary mixtures
 suggests the same mapping. The interaction of  both
 components with the surface  is  attractive   and the condition
 $\left|h_1\right|$ =  0 in the Ising model is fulfilled at  some nonzero
 liquid-surface interaction potential, which  is unique  for any  definite binary
 mixture \cite{Lawrev}.  So, binary mixtures belong to the
 universality class of normal transition  in the sense, that they show
 preferential adsorption  of one component  ($\left|h_1\right|$ $\neq$
 0).  Nevertheless,  experimental studies  show, that the behaviour  of the
 local order parameter, i.e. the difference between the concentrations
 of  the coexisting phases  near the  surface at  T $<$  T$_C$, follows
 a power law with the  critical exponent of an ordinary transition, at
 least    in   a   case  of a weak preferential   adsorption
 \cite{Fenzl,Franck}. The  crossover   from   ordinary  to   normal
 transition, predicted for the Ising model \cite{Ritschel}, was confirmed
 experimentally for supercritical binary mixtures  \cite{Law} and
  a crossover distance up  to dozens of nanometers was  obtained for weak
 preferential  adsorption.
\par  The surface  critical behaviour  in a
 one-component  fluid is  essentially  less clear  in comparison  with
 binary  mixtures.  Interaction with the surface and the effect of  missing
 neighbours near  the surface are physically  meaningful for real molecules.
 The only possible definition of the second component as being represented by the
 voids introduces a drastic  asymmetry of  the system, because  this "component"
 does not show  any interaction,  effect of  missing neighbours, etc.
 This makes it difficult to define properly the order parameter
 and  to find   the strength of the fluid-surface  attraction,   which   provides
  the condition  for the ordinary
 transition ($\left|h_1\right|$ =  0).   For example, this  condition may  be attributed  to the
 horizontal density profile of critical fluids \cite{Evans}. This could
 be    achieved    at  some definite    attractive    fluid-wall
 interaction.   Strengthening   of   fluid-wall   interaction   causes
 preferential adsorption  of molecules. Note, that in  this regime a
 crossover from ordinary to normal transition with the distance to the
 wall for  LJ fluid was  found \cite{Evans}.  To our  knowledge, there
 are no experimental  studies of the surface critical  exponent of the
 order parameter  in a one-component  fluid.  The only  simulation study
 was reported for  water near hydrophobic surface, where  the value of
 the exponent ${\beta_1}$ was found  close to Ising value for ordinary
 transition  \cite{BGOPCCP}.
\par  Surface  transitions as well as the
 surface critical behaviour are modified in the cases of more than one
 confining surface and (or) non-planar surfaces (for example: in the case of fluids
 in  pores or large solutes and  colloids in  fluids). The phase transition
 of fluid in pore is shifted with respect to the bulk one in $\mu$-T plane
 (see brief review in our recent paper \cite{BGO2004}).
 This distorts  the  surface phase transitions
 \cite{Dietrichrev,Evansprew}  as well  as the surface critical behaviour.
 The local critical behaviour  depends strongly on the particular shape
 of  the surface  (corners,  edges, cones,  cubes, parabolic  surfaces
 etc.)    both   in the cases of ordinary and normal transitions
 \cite{Cardy,Pleimling,Dietrichedge,Igloi}.  As a result, the critical
 surface exponent becomes non-universal,  i.e. dependent on the system
 shape.  In some cases, for example for parabolic surfaces, the surface
 critical behaviour is no longer  characterized by the usual power laws
 but by stretched exponentials \cite{Igloi}.
  \par   For
 experimental  and   computer  simulation  studies   of  the  surface
 transition and surface critical behaviour in one-component systems, the
 fluid   density   distribution   near   the  surface   is   the   key
 property. Experimental studies of the liquid density near solid surfaces
 show oscillations \cite{Dutta,Tolan}    and    depletion
 \cite{Chan,Tolan,thomas,fin1,hrun1,jensen,Pramana}.         Noticeable
 depletion of the liquid water density near  various hydrophobic surfaces
 was detected by  neutron reflectivity \cite{thomas,fin1,hrun1}, X-ray
 reflectivity  \cite{jensen} and electric  conductivity \cite{Pramana}
 measurements. Liquid water density of about 85 to 90 $\%$ of the bulk
 value was found in the interval of 20 to 50 $\mbox{\AA}$ from the surface by
 neutron  reflectivity  measurements \cite{thomas,fin1,hrun1}.   These
 experiments did  not allow  to get water  density profiles,  but they
 clearly indicate the absence of  a vapour layer at a surface.  The
 density  profiles, obtained from  X-ray reflectivity  measurements of
 water at a paraffin surface  \cite{jensen}, show a water density of about
 90 to 93 $\%$  of the bulk value at distances  less then 15 $\mbox{\AA}$
 from the surface. A much  stronger density depletion, which appears as
 gradual density  decay and  not as formation  of a vapour  layer, was
 observed in the case of n-hexane at a silicon surface \cite{Tolan}.
 Note  also, that  adsorption  measurements show  strong depletion  of
 the liquid  neon density near an extremely  weakly attractive  cesium
 surface  when  approaching the neon  critical  temperature  \cite{Chan}, however they
do not allow to distinguish  between
a gradual  density  decay and the formation of a vapour  layer.
\par  The
 presence  of  nanobubbles  at  hydrophobic  surfaces  in  water  (see
 reference \cite{Attbubble} for a recent review)  may,  in  principle,
 indicate the formation of  a  water vapour  layer.   However, the amount  and
 size of the nanobubbles  decrease essentially  in degassed  water
 \cite{fin1}   and  increase  due   to  contact  with air
 \cite{Vinogr1}. Obviously, water + air mixtures rather then the liquid-gas
 coexistence of pure water should be considered in order to elucidate
 the  origin of  the  nanobubbles.
 \par  The experimentally  observed
 attraction between hydrophobic surfaces in liquid water \cite{Christ}
 could be related both to the shift of the phase transition due to the
 confinement  and   (or)  to   the  peculiarities of the water  density
 distribution near the hydrophobic surface. In equilibrium with
 a bulk  reservoir, a fluid  confined between walls  exists as  a vapour
 (capillary  evaporation)  or as  a  liquid (capillary  condensation),
 depending on  the strength of fluid-wall interaction,  bulk state and
 distance  H  between   the  walls  \cite{Evansrev}.   In  particular,
 the phenomenon  of capillary evaporation  corresponds to  the equilibrium
 between a bulk liquid  and confined vapour.  Capillary evaporation was
 obtained    in    computer    simulations    of   both    LJ    fluids
 \cite{Att9397,Evansevap,Henderevap}             and             water
 \cite{Brov0002,Bratko,Pertsin}.  The phase  state of a confined fluid
 in  equilibrium  with the bulk liquid under its equilibrium vapour pressure
  \cite{Evansevap,Brov0002}
 depends on the  strength of the fluid-wall interaction  only, and, in
 particular, for  weakly attractive walls only the vapour phase  is stable
 for any distances  H between the confining (infinite)  walls.  In the
 case of an oversaturated  bulk the capillary evaporation takes place
 at   some   critical    distance   H$_{crit}$   between   the   walls
 \cite{Att9397,Henderevap,Pertsin}, which depends on the magnitude of the
 applied bulk pressure.  A simple estimation of  the oversaturation of water at
 ambient conditions  \cite{Chandler99}, based on  the Kelvin equation,
 predicts  a value H$_{crit}$  of about  1000 $\mbox{\AA}$.   At  H $<$
 H$_{crit}$ the metastable liquid  approaches a liquid spinodal with
 decreasing H,  that  could produce  long-range attractive  forces
 between the  walls \cite{Att9397,Att92}.  Another  explanation of the
 hydrophobic  attraction is  based solely  on the  depletion  of water
 density near the hydrophobic surface  and this effect was studied for
 water between  hard walls  \cite{Forsman}.
\par The  above presented
 hydrophobic attraction is  only one of the phenomena,  which could be
 fully understood  only based  on the knowledge  of the  surface phase
 behaviour of  fluids.  This includes,  in particular, the location  of
 phase transitions and knowledge of the density distribution near the
 surface in various thermodynamic states.   In computer simulations
 surface  phase diagrams have to be  studied in slitlike
 pores, as periodic  boundary conditions are not  applicable to
 semi infinite system.  However, there are a limited number of simulated
 or experimentally determined coexistence  curves   of   fluids  in
 confinement  (see reference \cite{BGO2004} for  a  brief  review and  some
 recent studies \cite{Brod,Cummings,Binder04}).
\par In this paper we
 present coexistence  curves  of water  in  hydrophobic pores  of
 various sizes and shapes.  The main  goal of the present paper is
 studying the temperature dependence of the density profiles  of a fluid
 near a weakly attractive surface along the liquid-vapour coexistence
 curve and in supercritical states of the confined fluid. The obtained results are used for
 the analysis  of the surface  critical behaviour and  possible surface
 phase transitions.  Dimensional  crossover in pores and extrapolation
 of the observed surface critical behaviour to semi infinite systems
 are discussed.
\begin{figure}[ht]
\begin{center} \includegraphics [width=12cm]{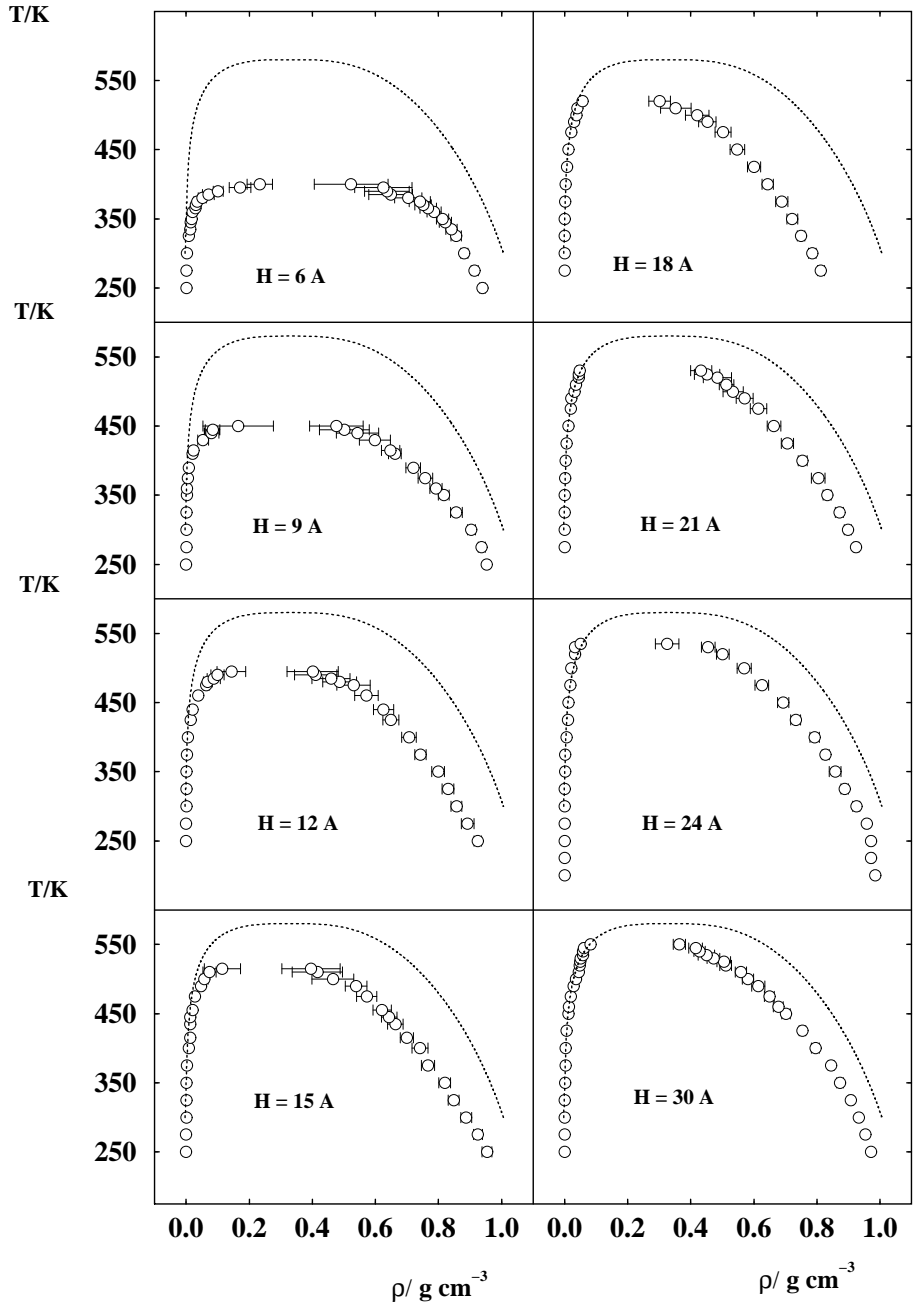}
\caption{Liquid-vapour  coexistence   curves  of  water   in  slitlike
hydrophobic pores.  Bulk coexistence curve is shown by dotted line.}
\end{center}
\end{figure}
\section{Method}
  TIP4P water  \cite{Jorgensen} was  simulated in slitlike  pores with
width H from 6 to 50 $\mbox{\AA}$ and in cylindrical pores with radius R
from 12  to  35 $\mbox{\AA}$.   A  spherical  cutoff  of 12  $\mbox{\AA}$
(oxygen-oxygen distance)  for both  the Coulombic  and  LJ parts
of the  water-water interaction potential was used. In accordance with the
original parametrization of the  TIP4P  model,  no  long-range
corrections  were  included.   The interaction between the water
molecules and the surface was described by a (9-3) LJ potential:
\begin{eqnarray}
U_{w-s}(r) = \epsilon [(\sigma/r)^9 - (\sigma/r)^3]
\end{eqnarray}
 where {\it r} is  the distance from the water oxygen to  the pore wall. The
parameter  $\sigma$ was  fixed  at 2.5  $\mbox{\AA}$, the
parameter $\epsilon$ = 1 kcal/mol (producing well-depth U$_0$ of  the
potential (3) of -0.385 kcal/mol) corresponds  to
a paraffin-like hydrophobic surface. The average  water
density  in  the pore  was calculated, assuming  that the water
occupies a pore volume  till the distance  $\sigma$/2 =  1.25
$\mbox{\AA}$  from the  pore  wall.  Coexistence curves of water
in pores in  a wide temperature range were obtained by Monte Carlo
simulations in  the Gibbs ensemble  \cite{Panagiot}.  For the three
largest pores (H = 50 $\mbox{\AA}$, R = 30 and 35
$\mbox{\AA}$) the  densities of  the coexisting  phases were
obtained only  for two temperatures  (300  K  and  520  K).  The
number of successful transfers per molecule between the coexisting
phases varied from 5 to about 100 in largest  and
smallest pores, respectively.  The number of water  molecules in
the two simulation  cells varied from  about 400 in the smallest pores up
to 8000 in  the largest cylindrical pore with R = 35 $\mbox{\AA}$.
The ratio L/H for slitlike pores or ratio L/2R for cylindrical
pores, where L is the longitudinal size of  the simulation cell,
always exceeded  1 and  at high  temperatures achieved  the value
10  in the smallest  slitlike  pores.  The  density  profiles  of
the  coexisting liquid and vapour phases and in some supercritical
states were obtained by subsequent Monte Carlo simulations in the NVT
ensemble.
\section{Results}
The  simulated coexistence  curves  of water  in hydrophobic
slitlike pores  of various  sizes are  shown in figure 1. The
liquid density of confined  water is essentially  lower then  in
the bulk in  the whole studied  temperature intervals.   The
lowering of the pore  critical temperature  and the apparent
flattening of the top  of  the coexistence curves with decreasing
pore size (from H = 30 $\mbox{\AA}$ to H  = 6 $\mbox{\AA}$) is noticeable.
A similar decrease of the  liquid water density is
observed in hydrophobic cylindrical pores with radii R = 12
$\mbox{\AA}$, 15 $\mbox{\AA}$,  20 $\mbox{\AA}$  \cite{BGO2004}
and R  = 25 $\mbox{\AA}$  (figure 2).  Note, that in the  case of
cylindrical pores at  some temperature (about 460 to 470 K  for
the pore with R = 25 $\mbox{\AA}$) the decrease  of the liquid density
becomes more pronounced, producing a noticeable change  of
the slope  of the  coexistence  curve diameter  (see figure 2).
In total 12 coexistence curves  of water in  hydrophobic pores - 8
slit-like pores (figure 1) and 4 cylindrical pores (figure 2 and
figure 12 of reference \cite{BGO2004}) - were  used in  this paper for
the analysis of various properties of the coexisting phases of
confined water.
\begin{table}
\caption{Estimated pore  critical temperature T$_C^{pore}$  (in K) and
pore  critical density $\rho_C^{pore}$  (in g  cm$^{-3}$) of TIP4P water in
hydrophobic pores.}
\begin{indented}
\item[]\begin{tabular}{@{}lll}     \br     \lineup     Pore     width,
$\mbox{\AA}$&T$_C^{pore}$&$\rho_C^{pore}$    \\
\br
&Cylindrical    pores&\\
\mr
R = 12&535.0$\pm$15.0&0.220$\pm$0.007\\
R = 15&535.0$\pm$10.0&0.172$\pm$0.007\\
R = 20&540.0$\pm$5.0&0.156$\pm$0.007\\
R = 25&555.0$\pm$5.0&0.150$\pm$0.007\\
\mr
&Slitlike  pores&\\
\mr
quasi-2D water &330.0$\pm$7.5&\\
\mr
H  = 6&402.5$\pm$2.5&0.358$\pm$0.007\\
H  = 9&452.5$\pm$2.5&0.294$\pm$0.007\\
H  = 12&497.5$\pm$2.5&0.262$\pm$0.007\\
H  = 15&520.0$\pm$5.0&0.246$\pm$0.007\\
H  = 18&525.0$\pm$5.0&0.235$\pm$0.007\\
H  = 21&535.0$\pm$5.0&0.241$\pm$0.007\\
H  = 24&540.0$\pm$5.0&0.229$\pm$0.007\\
H  = 30&555.0$\pm$5.0&0.227$\pm$0.007\\
\mr
bulk 3D water&580.2$\pm$2.5&0.330$\pm$0.003\\
\br
\end{tabular}
\end{indented}
\end{table}
\begin{figure} [ht]
\begin{center}
\includegraphics[width=10cm]{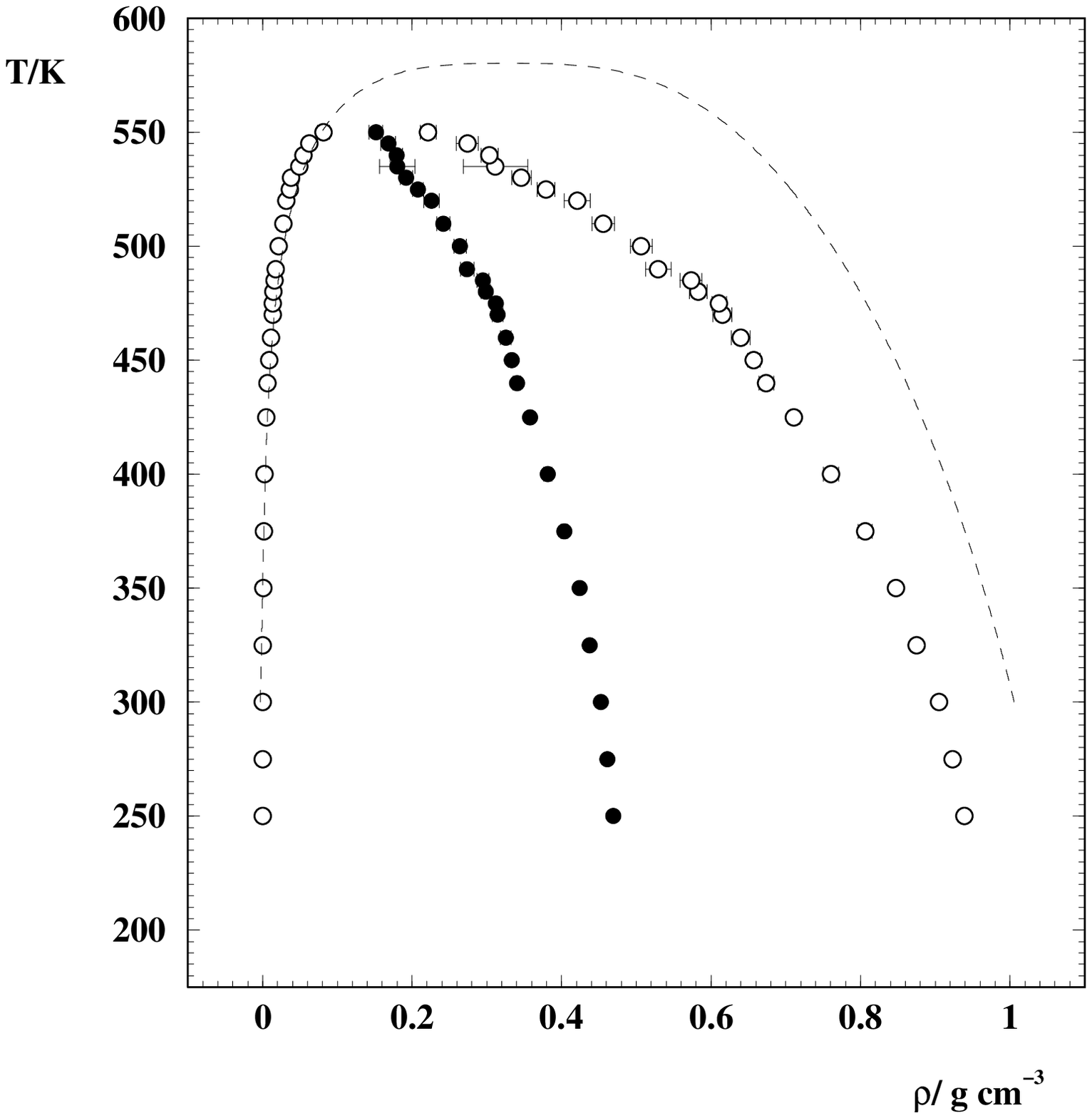}
\caption{Liquid-vapour  coexistence  curve  of  water  in the cylindrical
hydrophobic  pore   with  R  =  25  $\mbox{\AA}$.    Diameter  of  the
coexistence curve is shown  by closed circles.  Bulk coexistence curve
is shown by dashed line.}
\end{center}
\end{figure}
 \par We have estimated  the pore critical temperature T$_C^{pore}$ of
water  in various  pores as  an average  of two  temperatures:
the highest temperature, where two-phase coexistence  was obtained,
and the lowest temperature, where the two phases become identical in
the Gibbs ensemble MC simulations  (see \cite{BGO2004}  for
details).   The estimated values of T$_C^{pore}$
are presented in table 1. The largest   uncertainty  of
T$_C^{pore}$ is given  for  narrow cylindrical pores,
because the  decrease of the  sizes of the liquid and vapour domains
with  increasing temperature prevents an accurate location of  the
two-phase  region.   Additionally, we show in table 1 the
critical  temperature T$_{3D}$ of bulk  TIP4P water and
the critical temperature T$_{2D}$  of quasi-two-dimensional TIP4P
water with water oxygens  located in one  plane \cite{BGO2004}.
The evolution  of the pore  critical temperature  T$_C^{pore}$
with  pore size  is  shown in figure 3. The  shift of  the
critical temperature  in slitlike  pores was fitted to the power
law:
\begin{eqnarray}
\Delta T_C = (T_{3D} - T_C^{pore})/T_{3D} \sim (\Delta H)^{-\theta}.
\end{eqnarray}
\begin{figure}[ht]
\begin{center}
\includegraphics[width=10cm]{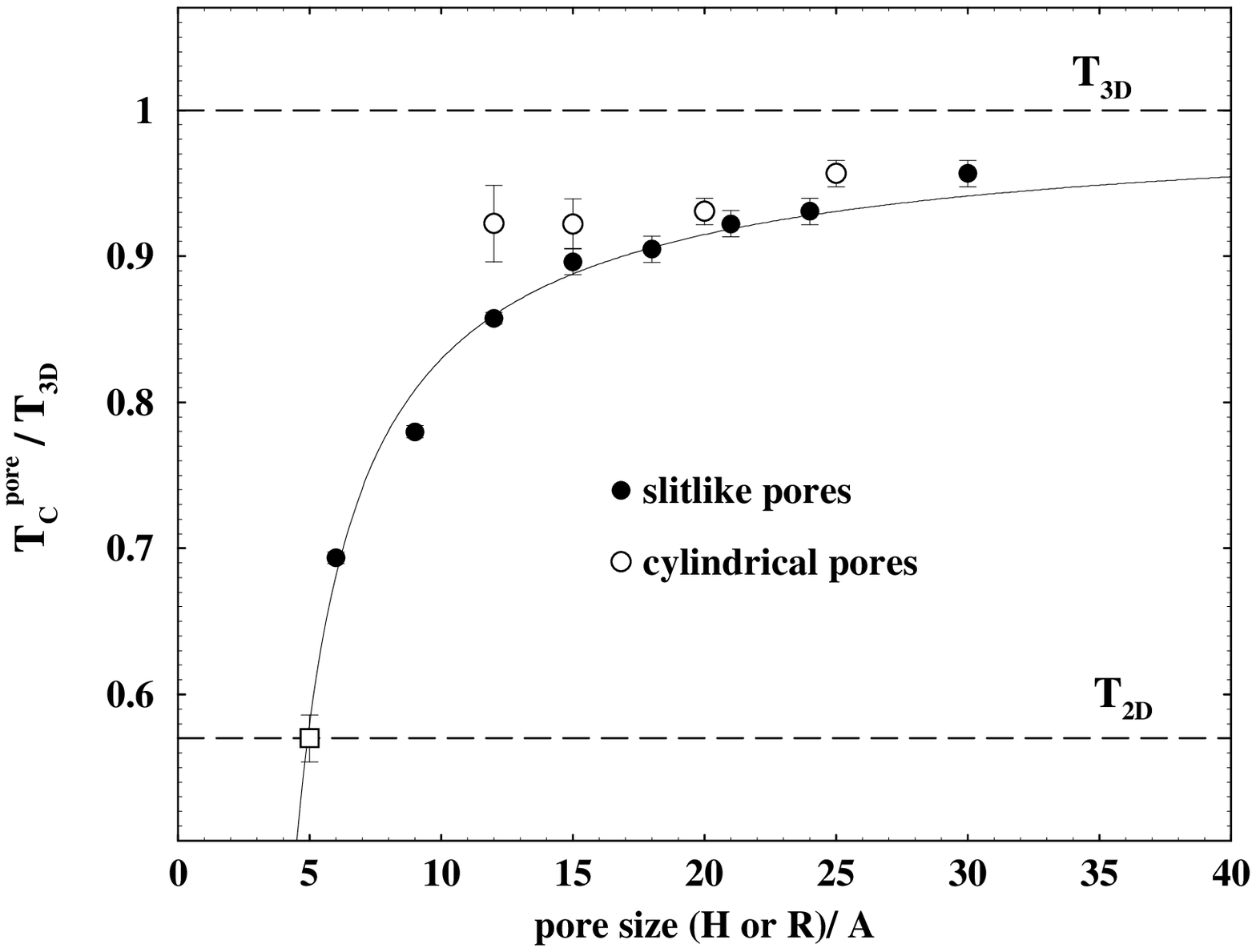}
\caption{Dependence of the pore critical temperature on the pore
size. Closed  symbols - slitlike pores. Open symbols  -
cylindrical pores. The critical  temperatures of bulk  and
quasi-two-dimensional water are  shown  by  dashed  lines. The square
corresponds  to  the  critical temperature of  the
quasi-two-dimensional water,  attributed to H  = 5 $\mbox{\AA}$.
The solid line is a fit of equation (4) to the data  for slitlike  pores (see text for details).}
\end{center}
\end{figure}
 In the  limit of a thick fluid film the value of
$\theta$ is expected to be equal to $\theta$ = 1/$\nu_{3D}$ 
($\nu_{3D}$ = 0.63 \cite{beta}) from finite-size  scaling   arguments
\cite{Fisher1}.  For thin film the effective value of $\theta$ should
be equal to 1 \cite{Evansprew,Evanslambda}. The thickness of the fluid
film $\Delta$H in a slitlike pore is  lower than the pore width H by
 $\sigma$  =  2.5   $\mbox{\AA}$,  the  LJ  parameter for
the water-substrate  interaction.    We  attributed  quasi-two-dimensional
water to water in a pore with effective size H = 5 $\mbox{\AA}$,
that  approximately  corresponds  to a  water monolayer,  strongly
localized  in one  plane.
\par  A fit of equation (4) to the the data for
slitlike pores (solid line in figure 3) gives the value $\theta$ =
0.82.
\begin{figure}[ht]
\begin{center}
\includegraphics[width=10cm]{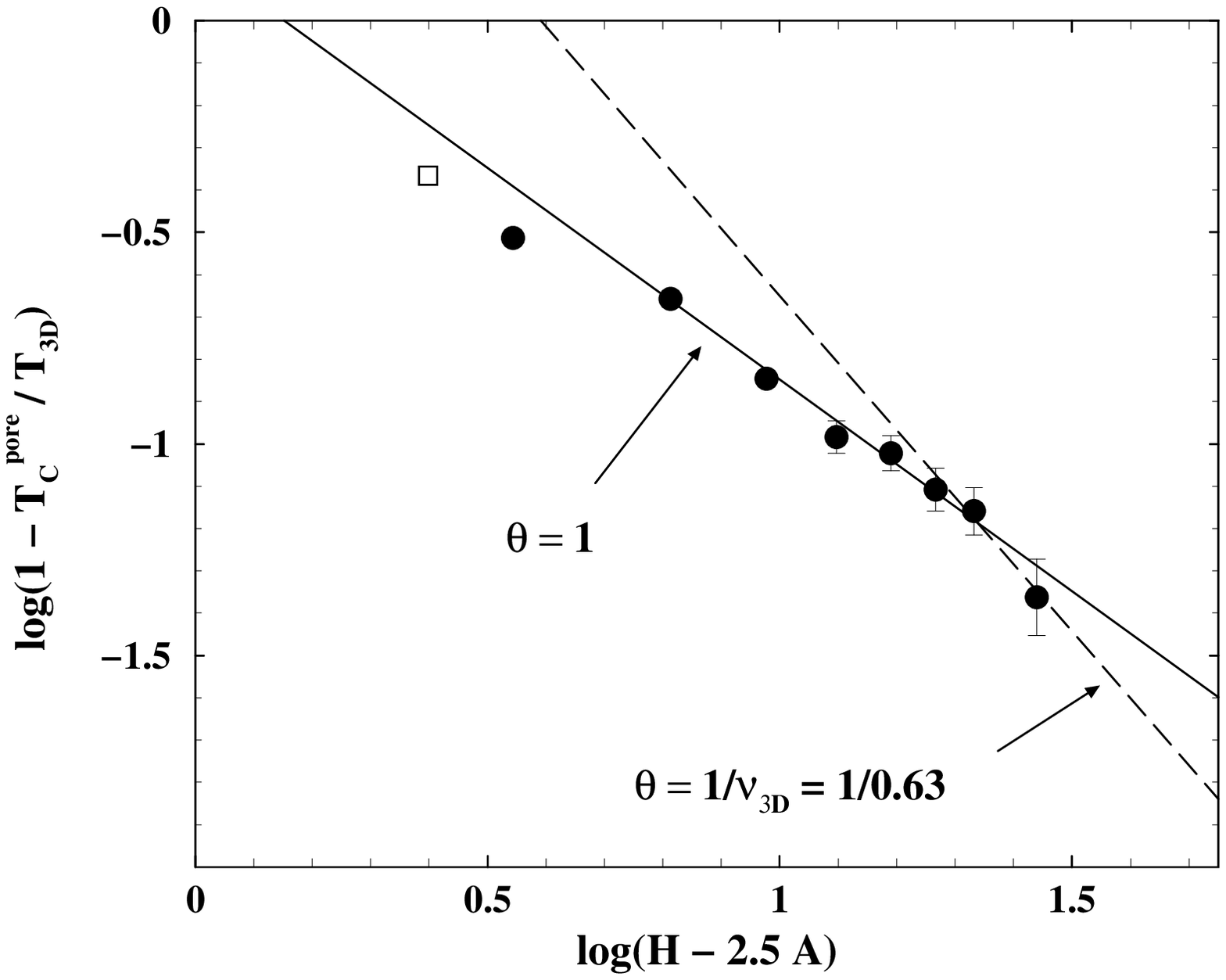}
\caption{Dependence of  the pore critical  temperature on the  size of
the slitlike pore  in double logarithmic scale. The square  corresponds to the
critical temperature of the quasi-two-dimensional water, attributed to
H = 5 $\mbox{\AA}$.}
\end{center}
\end{figure}
However, a plot of $\Delta$T${_C}$ vs
$\Delta$H in double logarithmic  scale (figure 4) shows most of
the data points may be well fitted by a linear law  ($\theta$ =
1), whereas the largest pore studied (H = 30 $\mbox{\AA}$) shows
a trend toward $\theta$ = 1/0.63. This suggests a crossover
between two  kinds of behaviour at a pore size H = 24
$\mbox{\AA}$, that corresponds to a film of roughly  8 molecular
diameters width.  Note,  that the  data points both for   the
narrowest   pore   (H   =   6   $\mbox{\AA}$)   and
quasi-two-dimensional  water  bend  down  from the  linear
dependence (figure 4).
\begin{figure}[ht]
\begin{center}
\includegraphics[width=10cm]{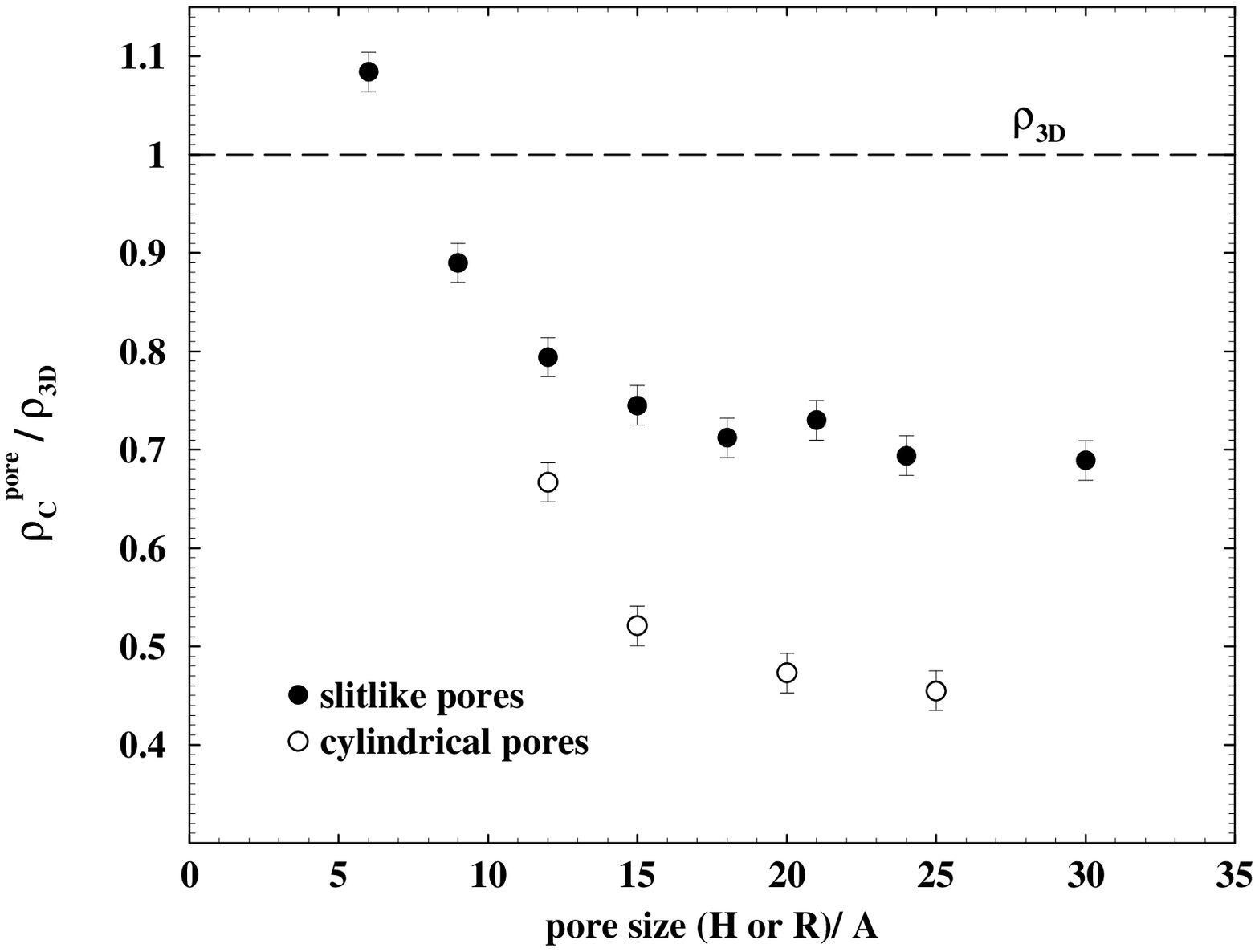}
\caption{Dependence of  the pore  critical density, normalized  to the
bulk critical  density, on the  pore size.  Closed symbols  - slitlike
pores.  Open  symbols - cylindrical  pores. The critical density  of bulk
water is shown by dashed line.}
\end{center}
\end{figure}

\par  A polynomial  extrapolation of  the  diameter  of the  coexistence
curves  $\rho_d$  =  ($\rho_l$  +  $\rho_v$)/2 to  the  pore
critical temperature  was   used  to  estimate  the   pore
critical  densities $\rho_C^{pore}$, presented  in table 1.   In
all pores,  excluding the smallest one, the pore critical  density
is essentially below the bulk value $\rho_{3D}$  = 0.330  g/cm$^3$
\cite{BGO2004}. Moreover,  in the considered  range  of   pore
sizes  (up  to  H  = 30  $\mbox{\AA}$) $\rho_C^{pore}$
decreases  with increasing  pore  size  and shows  no tendency towards
the bulk critical value (figure 5).
\begin{figure}[ht]
\begin{center}
\includegraphics[width=10cm]{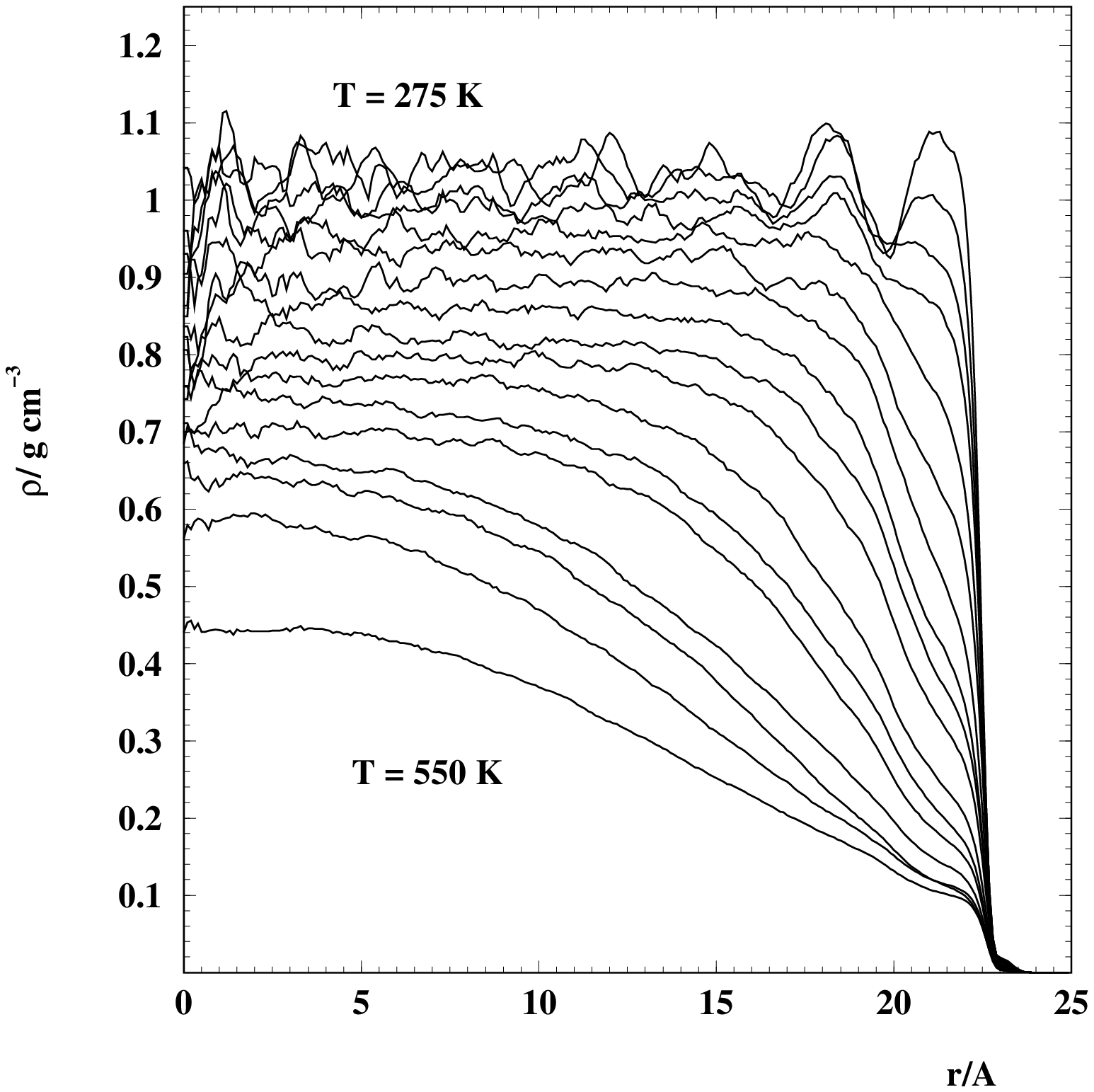}
\caption{Density profiles of water in the liquid  phase along the coexistence
curve in the cylindrical pore with R = 25 $\mbox{\AA}$.}

\end{center}
\end{figure}
\begin{figure}[ht]
\begin{center}
\includegraphics[width=10cm]{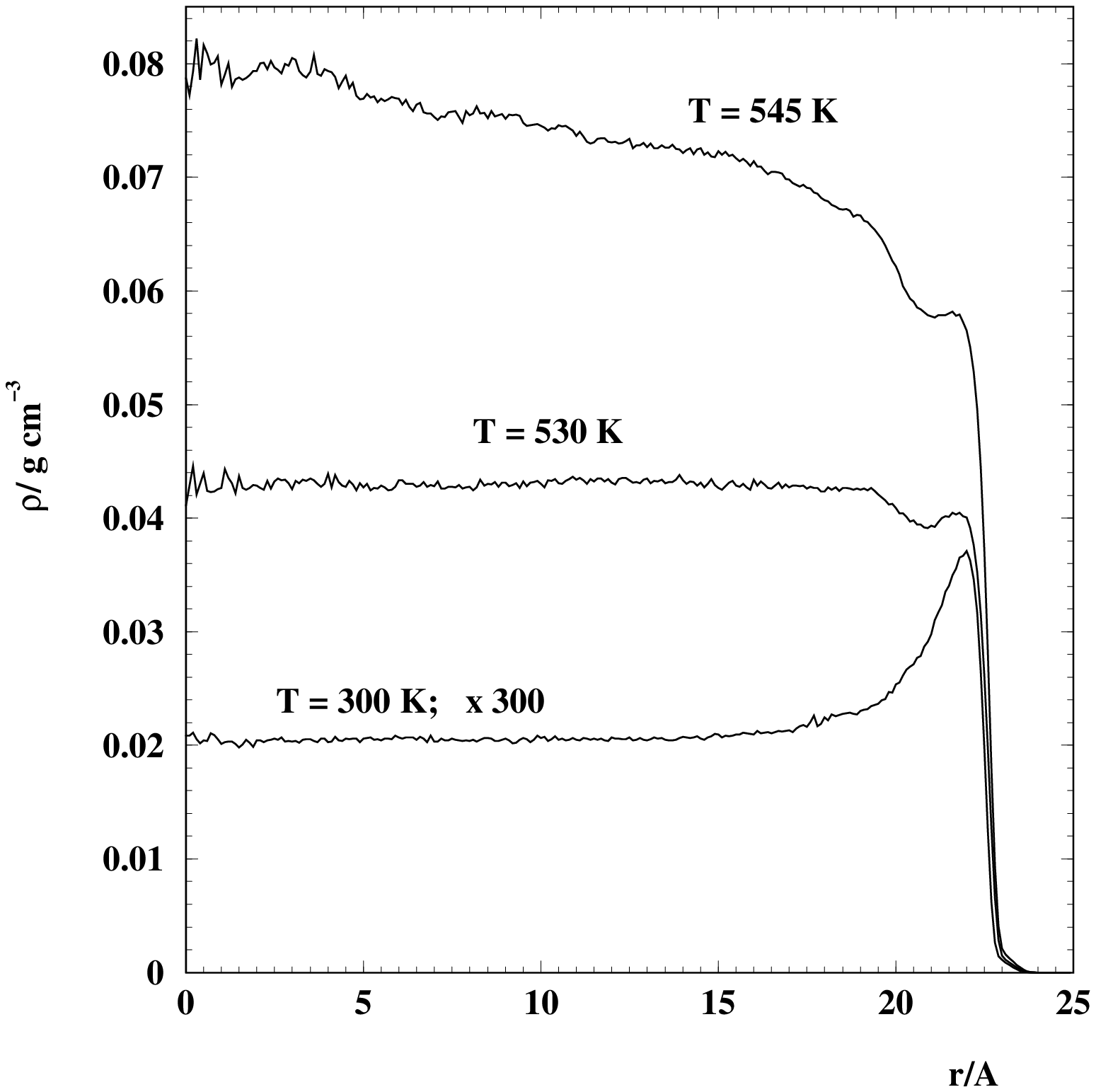}
\caption{Density  profiles of the water  vapour  phase along the coexistence
curve in the cylindrical pore with R = 25 $\mbox{\AA}$. For 300 K the density is multiplied by the factor 300.}
\end{center}
\end{figure}
\par  In  order  to  understand  the  observed  peculiarities  of  the
coexistence curves and to study the surface critical behaviour of
water in  hydrophobic  pores,  we  analyzed  the  density
profiles  of  the coexisting  phases  and  their  evolution with
temperature. Typical density profiles of water  in the liquid and
vapour phases are  shown in figures 6 and 7, respectively. The density
of the liquid phase  is strongly depleted  towards the  pore wall  and
this  depletion progressively intensifies with temperature
(figure 6) and pore  size.   This effect  is essentially stronger
in cylindrical  pores in comparison with slitlike pores.  Note, that the
depletion of the
density is accompanied by  an essential increase of
diffusivity   near  the  hydrophobic  surface  \cite{Broveps}.
\par The density profiles in  the vapour phase  show preferential
adsorption  of the water  molecules  at  weakly attractive  wall  at
temperatures below  $\approx$  500 K.   At higher temperatures the
vapour  phase shows a depletion  of the density  towards  the pore
wall similar to the liquid phase (figure 7). This means that at T
$<$ 500 K the density profile of the liquid bends downwards, whereas the
density profile of the vapour bends  upwards  when approaching the pore
wall. A similar behaviour was observed in lattice gas simulations
at temperatures below the wetting
transition \cite{BinderLandau} and in MC simulations of  LJ
fluids near a weakly  attractive substrate, including temperature close
to the critical temperature \cite{Colle}.
\par
Due  to the  spatial heterogeneity  of fluids in  pores  the
local coexistence curves,  i.e. the temperature dependence  of the
densities of the coexisting phases at various distances from the
surface, should be analyzed. Defining layers (and,
first of all, the surface layer) in  continuous models, which allows
a direct comparison with the results  for lattice  models, is  not
clear a priori. Far  from the surface, where  the packing
of the fluid molecules at the surface is negligible and the density
varies smoothly, the layer  thickness could be arbitrarily small.
Near the surface, both the packing  effect and the details   of
the fluid-surface  interaction,   become   important. Therefore,
it seems reasonable  to consider the local densities $\rho_i$,
averaged  over layers  of one  molecular diameter  thickness.
The location of  the first  (surface) layer was  chosen between
the first minimum  in  the  liquid  density  distribution  at  low
temperatures (typically  about   5  $\mbox{\AA}$  from  the  pore
wall)  and  the van der Waals water-wall contact  ($\sigma$/2 =
1.25 $\mbox{\AA}$ from the wall).  The thickness of the subsequent
water layers was set to 3 $\mbox{\AA}$. The average density of the
"inner" water near  the pore center was calculated  for a
layer extending 4 to 8 $\mbox{\AA}$ from the pore center.  The
coexistence curves of water in the surface layer and of water in the
pore interior for some cylindrical and slitlike pores  are shown
in figure 8.   While the  coexistence curve  of "inner" water
(figure 8,b) is  close to the bulk one (except for the proximity of
the pore critical temperature),  the surface water  shows
a drastically different behaviour  (figure 8,a). The density of the liquid
phase in the surface layer approaches extremely low values
(essentially below $\rho_{3D}$) with   increasing temperature and
the coexistence curve  shows triangularlike shape.  Note the highly
universal behaviour  of the "inner" water and the rather similar behaviour
of the surface water  in the  pores of various sizes and shapes.
\begin{figure}[ht]
\begin{center}
\includegraphics[width=10cm]{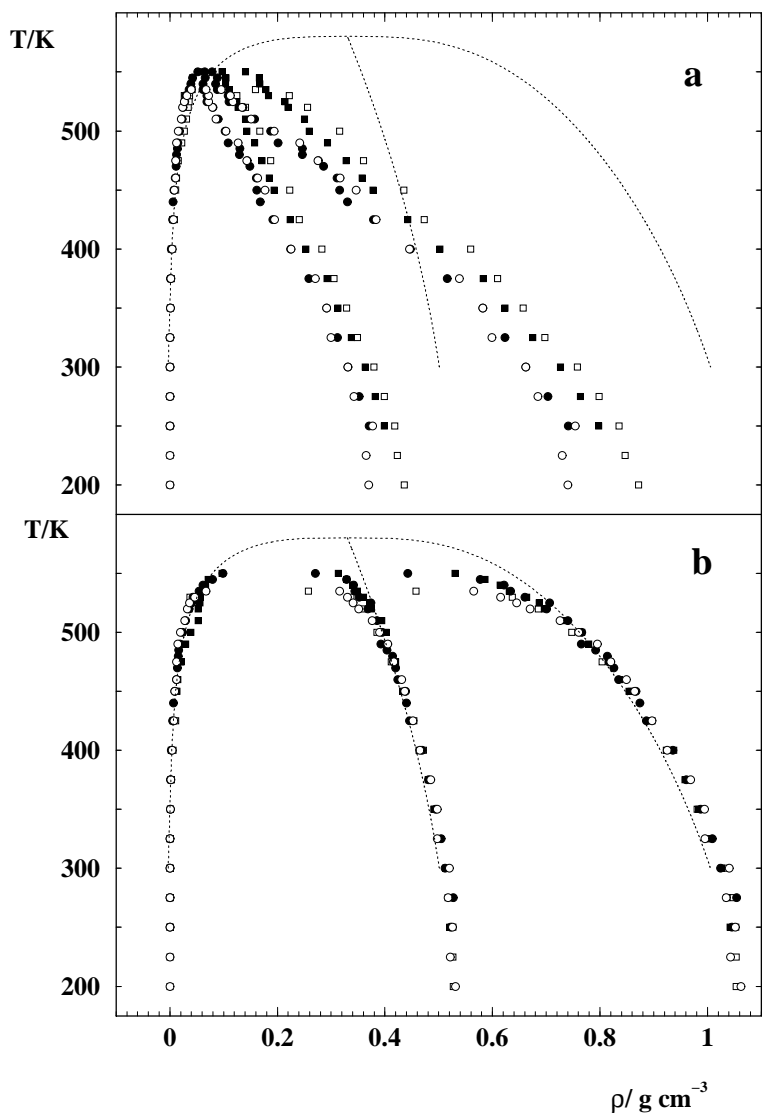}
\caption{Coexistence  curves and with diameters of water  in  hydrophobic pores:  closed
circles -  cylindrical pore with R  = 25 $\mbox{\AA}$;  open circles -
cylindrical pore with  R = 20 $\mbox{\AA}$; closed  squares - slitlike
pore with H  = 30 $\mbox{\AA}$; open squares - slitlike  pore with H =
24  $\mbox{\AA}$.  a)  water in the surface  layer;  b)  water  in the pore
interior. Coexistence  curve and diameter  of bulk TIP4P water are  shown by
dotted lines.}
\end{center}
\end{figure}
\par  The analysis of  the shape  of the  coexistence curves  in fluid
systems includes  the temperature dependence of  the order parameter
and diameter.  Because of the zero critical magnetization of an Ising system
its  order parameter  is  simply the magnetization  (bulk  or local).   In
confined fluids (as well as in the bulk)  the critical density $\rho_C$
is  non-zero  and in the asymptotic  limit  the  order parameter  is  the
deviation of the densities $\rho_l$ and $\rho_v$ of the coexisting liquid and vapour phases from  $\rho_C$.  Apart from  the critical
point the temperature dependence  of the diameter $\rho_d$ = ($\rho_l$
+ $\rho_v$)/2 should be taken  into account and so the order parameter
is defined as the deviation of the density from the diameter $\Delta\rho$  = ($\rho_l$  - $\rho_v$)/2.   We
define  the  local  order parameter  of  a  fluid  in a  similar  way:
$\Delta\rho$(z)= ($\rho_l$(z)  - $\rho_v$(z))/2, where the densities of
liquid and vapour are taken at the same distance z  from  the
surface. The temperature dependence of  the bulk order parameter  near the
critical  point in magnets,  as well  as in  fluids, obeys  the simple
scaling law:
\begin{eqnarray}
  \Delta\rho \sim \tau^{\beta},
\end{eqnarray}
 where $\tau$  =($T_{3D}$ - T)/$T_{3D}$  is a reduced  temperature and
$\beta$ =  0.326 \cite{beta}  is a universal  critical exponent of
3D (bulk) systems. The local order parameter $\Delta\rho$(z) should obey the
same law, but  with another  critical  exponent, which  depends
on the  surface universality  class. The temperature dependence  of  the
local order  parameter $\Delta\rho_i$  for surface water, inner water and two
intermediate
water layers are shown in figure 9 in double logarithmic
scale. The slope of these curves is equal  to  the
exponent $\beta$ of the  power  law, equation (5). The  order parameter of
"inner" water follows closely the  bulk behaviour  up to $\tau$ =
0.08. The order  parameter  in  the surface  layer  shows
an essentially different  behaviour:   starting  from   extremely
low temperatures  $\tau$ = 0.57  (close to  freezing temperature
$\tau$ = 0.59  of  TIP4P  bulk  water \cite{TIP4Tfreez})  the
order  parameter follows  the scaling  law with
value  of the  exponent $\beta$  close  to the  value $\beta$$_1$
=  0.82 of  the  ordinary transition  in the  Ising magnets.   The
intermediate  two  layers  show a crossover from bulk like behaviour
to surface behaviour at $\tau$ = 0.22 and $\tau$  =   0.11,  in
the   second  and   the  third   layers, respectively. A higher
value  of the  exponent $\beta$  in  the surface layer  means,
that  the densities  of the  coexisting phases  near the surface
approach each other much  faster with temperature, than in the
pore interior.
\begin{figure}[ht]
\begin{center}
\includegraphics[width=10cm]{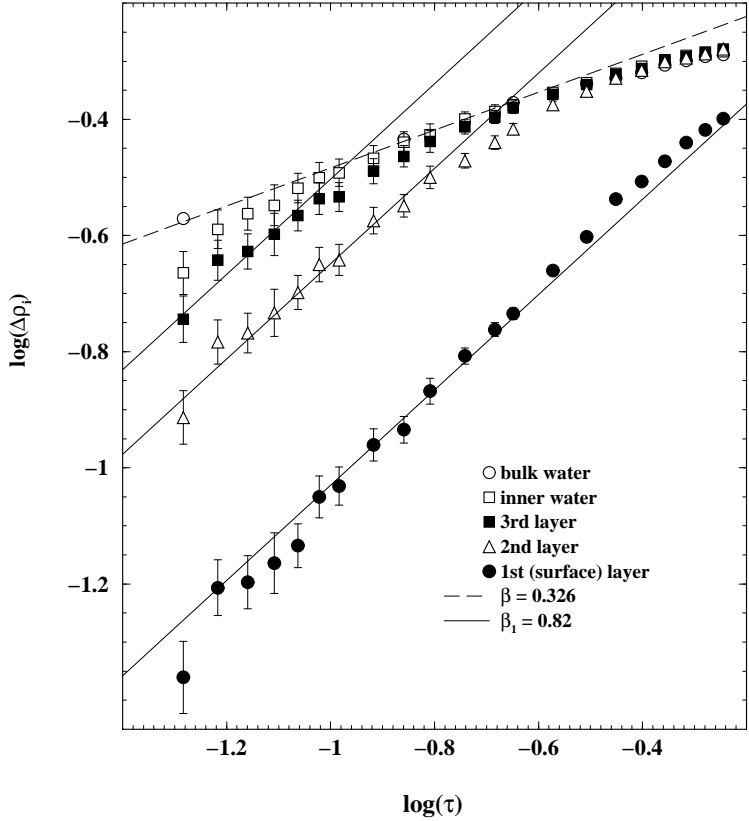}
\caption{Variation  of the local  order parameter  $\Delta\rho_i$ with
reduced temperature $\tau$  in the slitlike pore with H  = 30 $\mbox{\AA}$
 (double logarithmic scale).}
\end{center}
\end{figure}
\par This  disordering effect  of the surface  may also be illustrated,
using an effective exponent $\beta^{eff}$, defined as:
\begin{eqnarray}
\beta^{eff}(z,\tau) = d{\it ln}(\Delta\rho(z,\tau))/d{\it ln}\tau.
\end{eqnarray}
The value of  the exponent $\beta^{eff}$ is close  to the
value of some true critical exponent in the temperature intervals, which are outside the crossover regions. $\beta^{eff}$(z), obtained
by applying equation (6) to $\Delta\rho(z, \tau)$ for the interval
$\tau$ $\leq$ 0.31, is shown in figure 10 as a function  of the
distance to the pore center.   In  fact, the  value $\beta^{eff}$
is close to the 3D  Ising value in the pore interior and varies
between 0.8 and 1.0  in the  surface layer  (hatched area in
figure 10).  Note, that the maximum of $\beta^{eff}$ corresponds to
the minimum of the LJ water-wall  potential, which causes a slight
hump of the liquid density profile (see figure 10).
\begin{figure}[ht]
\begin{center}
\includegraphics[width=10cm]{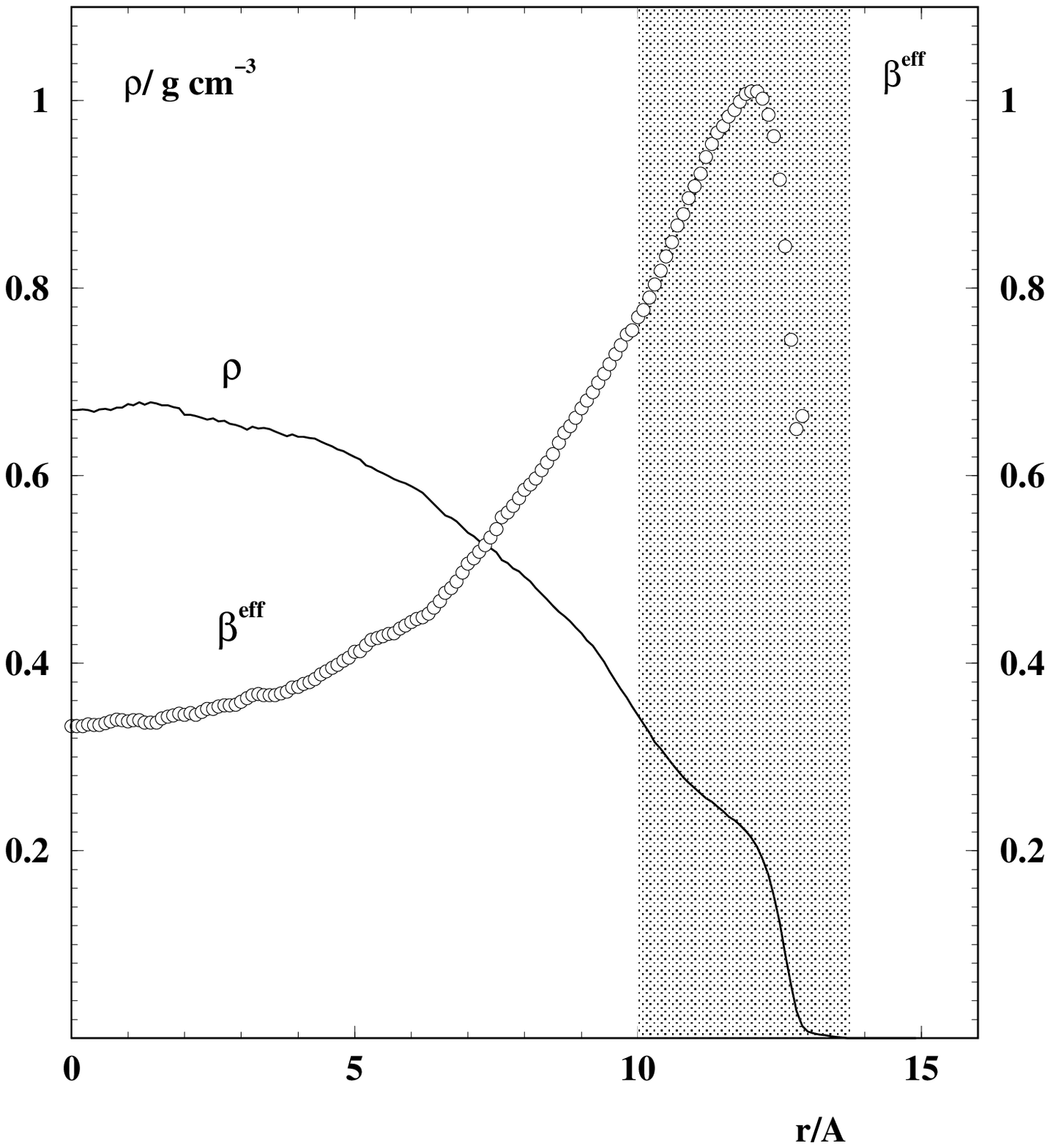}
\caption{Variation of the effective exponent  $\beta^{eff}$,
obtained by the applying equation (6) to the local order parameter
$\Delta\rho(z, \tau)$, with approaching pore  wall in
the slitlike pore with  H = 30 $\mbox{\AA}$ (circles).  The liquid density
profile  at T  = 400 K  is shown  by a line. The hatched area indicates
the location of the first (surface) water layer.}
\end{center}
\end{figure}
\par  The behaviour  of the  order parameter  in the  surface  layer is
rather universal in the three largest slitlike pores  and is
close to the simple  power law with $\beta_1$  of about 0.8
(figure 11,a). In the pores   with H $<$ 21   $\mbox{\AA}$
the dependence {\it ln}($\Delta\rho_1$) vs {\it ln}($\tau$)
progressively  bends down at high temperatures (figure 11,b) preventing the location
of a proper temperature interval, where the simple
power law is valid.

\begin{figure}[ht]
\begin{center}
\includegraphics[width=10cm]{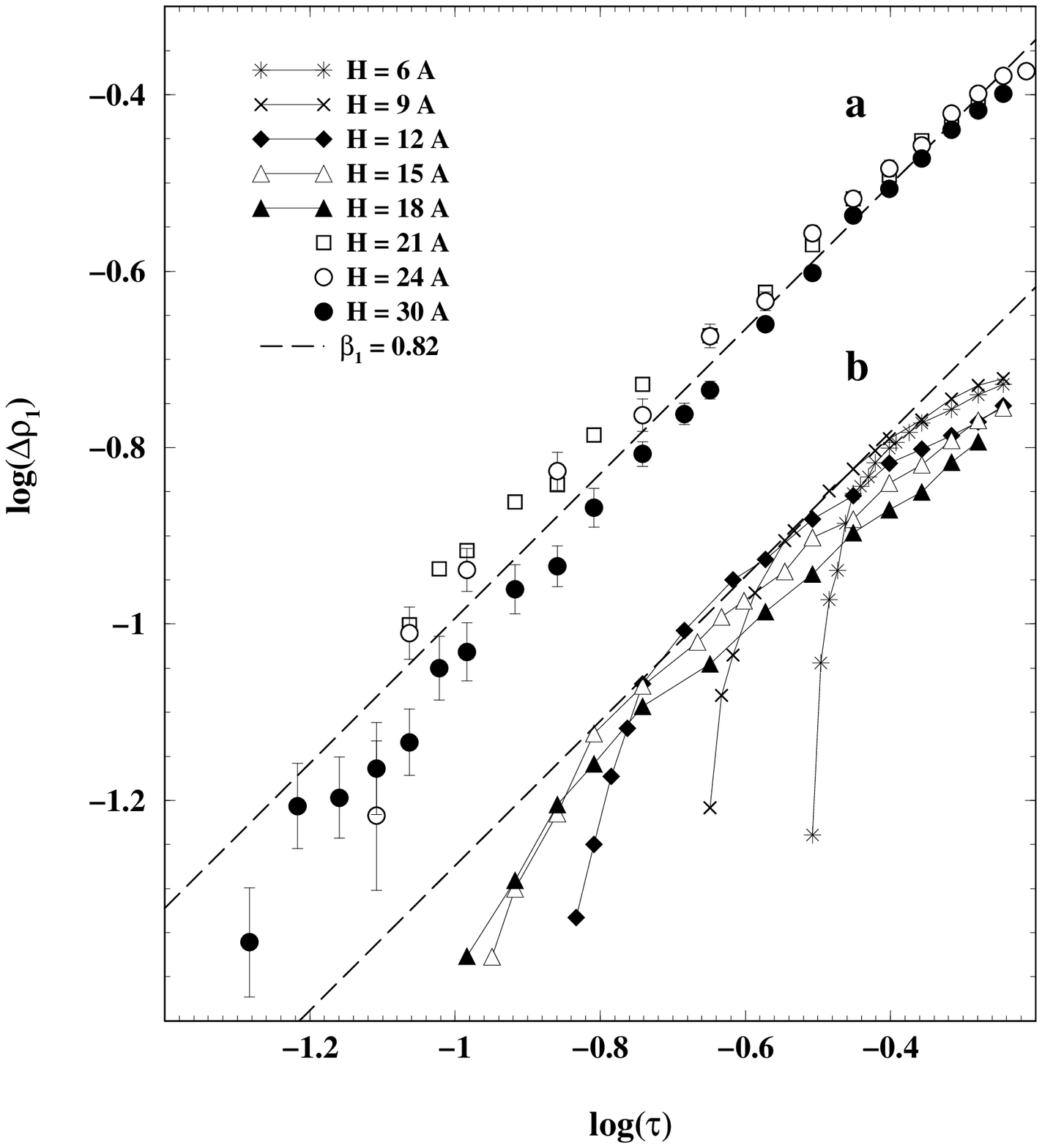}
\caption{Double logarithmic plot of the local order parameter in  the surface layer
$\Delta\rho_1$ vs.  reduced temperature  $\tau$ in {\it slitlike}  pores in
double logarithmic scale. (For the two narrowest pores all molecules were
attributed to  the surface  layer.) Data points  for the pores  with H
$\leq$ 18 $\mbox{\AA}$ are shifted by -0.40 (curves b).}
\end{center}
\end{figure}

\par Whereas the behaviour of water in the surface layer of cylindrical
pores  in general  looks  rather  similar to  that  in slitlike
pores (figure 8,a), the variation of  the local order parameter
$\Delta\rho_1$ with reduced temperature shows serious differences  (full circles in
figure 12).
$\Delta\rho_1$  follows roughly  the simple  power law  with
$\beta_1$ $\sim$ 0.8 at $\tau$ $\geq$ 0.20, whereas at  higher
temperatures a  simple power  law with another exponent of
$\beta_1$ of about 1.8 is  observed (figure 12).  In the second
and the subsequent  layers the behaviour with  the exponent
$\beta_1$ $\sim$ 0.8 could not be detected, but an apparent
crossover from the bulk-like  behaviour ($\beta$ $\sim$ 0.326) at
low temperatures to the
behaviour with $\beta_1$  of  about 1.8  could  be seen.   In the
second layer the bulk-like behaviour is  observed up to $\tau$
$\sim$ 0.18, whereas the behaviour with $\beta_1$ of about 1.8 is
valid at $\tau$ $\leq$ 0.11. In the  subsequent layers this
crossover region becomes  narrower and shifts to higher
temperatures.
\begin{figure}[ht]
\begin{center}
\includegraphics[width=10cm]{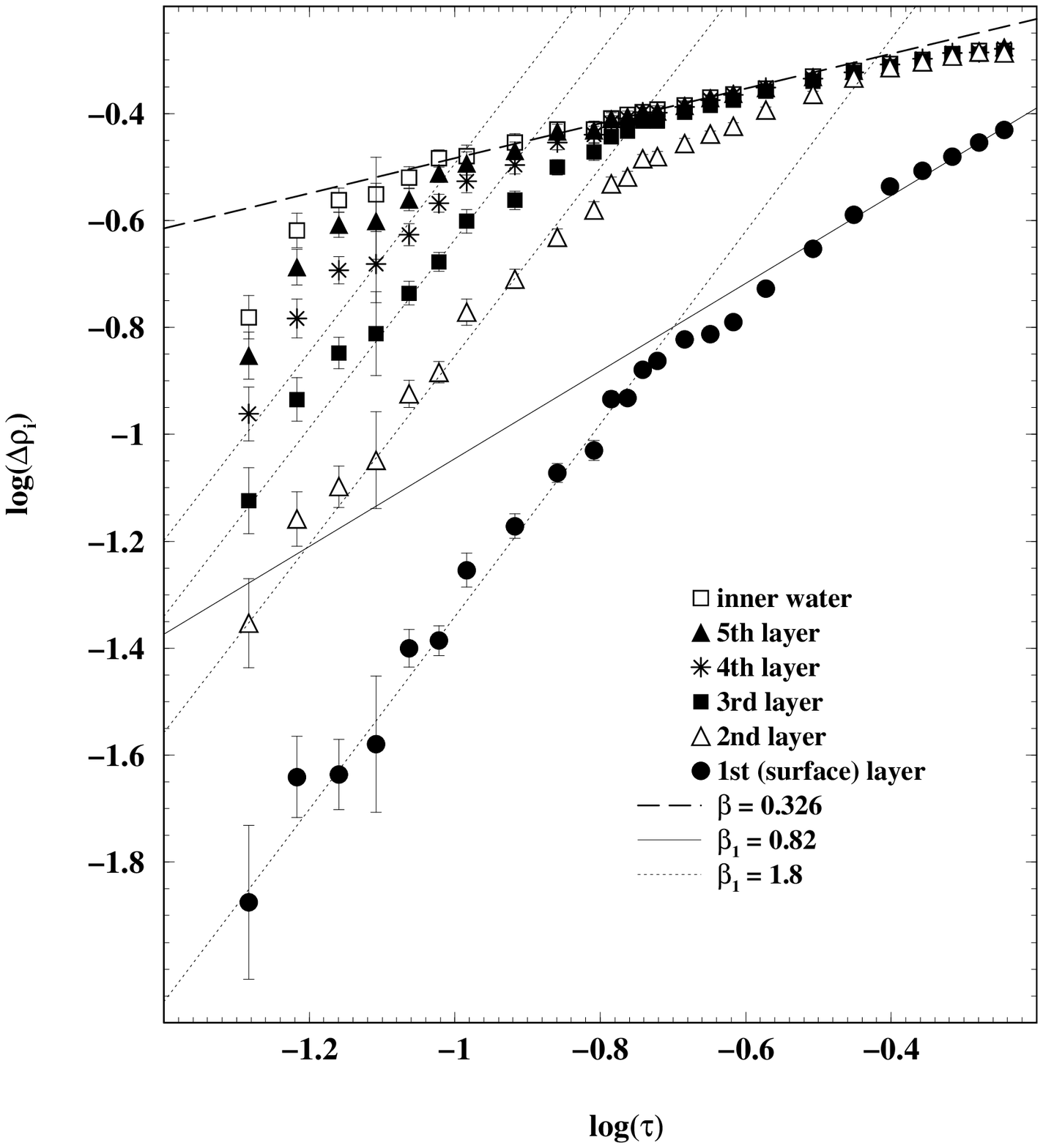}
\caption{Variation  of the local  order parameter  $\Delta\rho_i$ with
reduced  temperature   $\tau$  in a {\it cylindrical}   pore  with  R   =  25
$\mbox{\AA}$ in double logarithmic scale.}
\end{center}
\end{figure}

\section{Discussion}
 The  presented coexistence  curves and density  profiles give  us the possibility
 to  verify   theoretical  predictions  of confinement effects.  But they give us
also the possibility  to study phase transitions near single
surfaces by extrapolating
the pore results  to semi-infinite systems.
\par The  observed evolution of
the pore critical  temperature with the size of the  slitlike pores
is in general agreement with theoretical predictions and
simulations for the 3D Ising  model.  In  particular, the  shift
of  the  critical temperature in the  pores, which contain from 2
to 7 molecular layers is  inversely proportional  to the  pore
size (see figure 4 and equation (4)),  while in  the largest pore  studied  (9
molecular layers) a  crossover  to a power law  dependence with $\theta$
= 1/$\nu_{3D}$ is indicated.  In Ising films such a crossover  begins,
when their  thickness  achieves  4  to  8  layers
\cite{FisherN,Binder74,Schilbe96,Binder01}.  The critical
temperatures of quasi-two-dimensional water  and water  in a
pore, comprising single  water  layer (H  =  6  $\mbox{\AA}$),
deviate from  the  main dependence  $\Delta$T${_C}$ $\sim$
$\Delta$H$^{-1}$.   This deviation is in agreement  with the
behaviour  of  a  monolayer Ising  film
\cite{FisherN,Binder74,Schilbe96,Binder01}.   Obviously,   with
the decrease of the pore width from 2 to 1 monolayer, the system
loses any features  of three-dimensionality and becomes
essentially two-dimensional.
\par  The dimensional  crossover  from 3D  critical
behaviour to 2D behaviour in slitlike pores is expected at
temperatures,  where  the correlation length becomes comparable
with the pore size. It appears, particularly as  a  change of
the  critical exponent  of the  order parameter in the power law,
equation (5), where $\tau_{pore}$ = (T$_C^{pore}$ -
T)/T$_C^{pore}$ is used instead  of  $\tau$.  The  average
order parameter (i.e., the difference of the densities, which are  averaged  over the whole
pore) should be used here. In the four largest pores a
crossover to the 2D behaviour is  seen (figure 13,  left panel),
when the  correlation  length exceeds about 15 $\%$ of the pore
width (estimates of the correlation length are  given
below).
\begin{figure}[ht]
\begin{center}
\includegraphics[width=10cm]{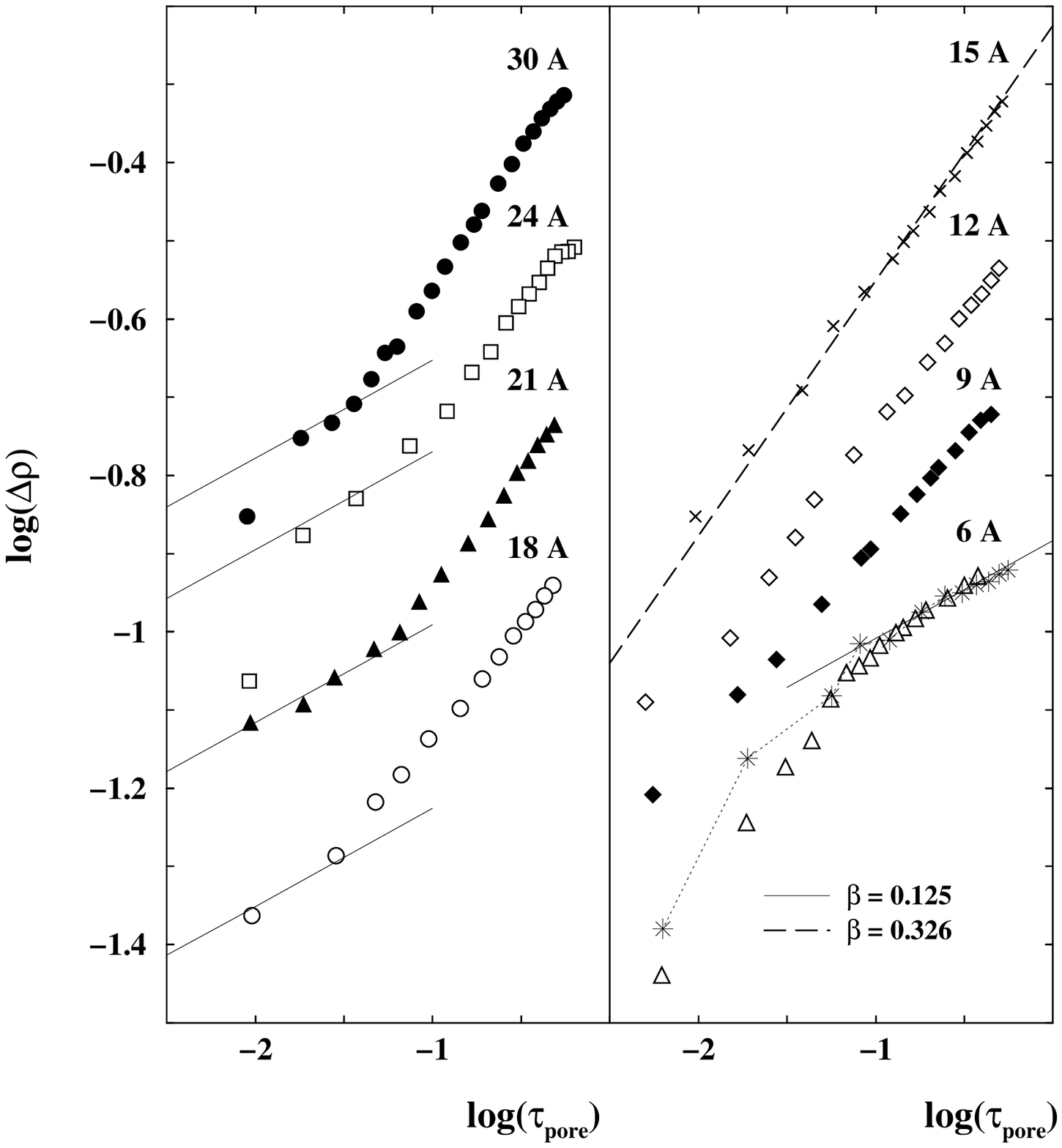}
\caption{Variation of the  order parameter $\Delta\rho$ from pore averaged
densities  with reduced  temperature $\tau_{pore}$  in slitlike
hydrophobic  pores   (double logarithmic  scale).  Pore  widths  H are
indicated  in the figure.   Data points  are shifted  progressively by
-0.20. The data points for the  layering transition are shown by stars
(see text for details).}
\end{center}
\end{figure}

\par A quite different  behaviour of the order parameter  is observed for
essentially  two-dimensional phase transitions, i.e. for
the liquid-vapour phase transition of water in the smallest studied
hydrophobic slitelike pore (H = 6 $\mbox{\AA}$, monolayer width)
and for the layering transition of water at a hydrophilic substrate
(slitlike pore with H =  24 $\mbox{\AA}$ and $U_0$ = -4.62
kcal/mol, \cite{BGO2004}).  In both cases the centers of water
molecules  are located in a  layer of about 1  to 2 $\mbox{\AA}$
width, the  critical temperatures of  both transitions are  close
(400 $\pm$ 5 K) and the order  parameter follows a power law with the 2D
exponent $\beta$ = 0.125  in the low temperature region (figure
13, right panel). When the  temperature exceeds about 375 to 385 K
($\tau_{pore}$ $\approx$ 0.04 to 0.06)
the  behaviour of the order parameter  in these systems
indicates a crossover  to a mean-field behaviour with exponent
$\beta$ = 0.5.  Such a crossover is  expected,  when the
correlation  length  becomes comparable  to the lateral size L of the simulation cell
\cite{BinderMons, Gubbins}.  In these ``one-layer systems'' the phase
transition  is  essentially   two-dimensional  and  so  the  2D
correlation  length diverges  at their  critical  temperatures.
A rough estimate with $\nu_{2D}$  = 1 and the 3D bulk value  of the amplitude
$\xi_{0}$ \cite{Bonetti}  shows a crossover to the mean-field
behaviour,  when  the 2D correlation length has grown to about 15 to 20$\%$
of L.  The lateral system size
used  for the  simulations of  a layering  transition  (\cite{BGO2004})is  about
50$\%$ larger than in  the case of the  hydrophobic pore with  H = 6
$\mbox{\AA}$. This causes a larger finite-size effect  in the latter
case.  Note, that a variation of the critical  temperature within
the interval estimated from GEMC (see table 1) does
not restore the asymptotic 2D behaviour in the hydrophobic
pore.
\par  In  the  pores  of
intermediate sizes  (H = 9,  12 and 15  $\mbox{\AA}$, i.e from 2
to 4 molecular  layers width) we  can not  observe any  tendency
to  the 2D behaviour, contrary to the expectation that  the  region  of  2D behaviour  should be essentially wider than in larger pores. This
can not be the result of a crossover to a mean-field behaviour due
to finite L,  as in the case of quasi-two-dimensional systems
considered above.  In the two largest pores (H = 24 and 30
$\mbox{\AA}$), where the lateral size L is about twice  the
pore size  H,  the  trend  to mean-field  behaviour  is noticeable
for  the highest temperature point only.   In smaller pores the
ratio L/H  is larger: from  3 to 10,  and therefore, no trend  to
mean-field behaviour should be expected even at  higher
temperatures than  in  the  largest  pores. The  apparent 3D  Ising
behaviour  in  a wide temperature interval, observed in pores of 2 to
4 molecular layer width (figure 13, H = 9 to 15 $\mbox{\AA}$), could be the result of
a competition between the trend towards 2D criticality and
a progressive contribution to the  order parameter from the  surface
layers, which show  faster disordering  compared  to the bulk.
Dimensional crossover  in  small  pores certainly
needs further studies  both for  fluid and lattice \cite{Binder01}
systems.
\par
The common feature of the coexistence curves of water in all
studied 12  hydrophobic pores is a significant decrease of the
density at the liquid branch in comparison  with the  bulk (see figures 1,2 and
also figure 12  in \cite{BGO2004}). An inspection of the  density
profiles evidences that this  decrease is due to the lowering  of the
density near the pore wall (figure 6), which appears as gradual decline
 of the water  density towards the surface  without any evidence for the
 formation of a vapour layer.  This agrees with the recent
experimental studies  of water density depletion  near
hydrophobic surfaces at  ambient  \cite{thomas,fin1,hrun1,jensen}
and  elevated \cite{jensen,C60} temperatures.
 The absence  of a vapour layer near
the  hydrophobic surface means  the absence  of drying  (or
predryring) transitions  at   T  $<$  T$_C$.  This  agrees   with
expectations from  theory, concerning the  phase behaviour  near  a
surface  with long  range fluid-surface  interaction
\cite{EbnerSaam,  Indekeu85} and is confirmed by computer simulations of LJ fluids
near weakly attractive walls \cite{Colle}. Our simulations of
water near an  extremely hydrophobic surface (less  attractive
than a paraffin-like  surface) suggest,  that a stable vapour layer
could never occur near a real  hydrophobic surfaces, which always
interacts  with water via  attractive long-range van der Waals
forces.   The possible  existence  of a  metastable vapour  layer  near a
 hydrophobic surface deserves  further investigations.
\par Due to the absence of
a surface transition, a study of  the surface critical behaviour of
the fluid becomes possible. The obtained density profiles  of the
coexisting phases  at various temperatures allow to study  the
temperature dependence  of the  order parameters  in various
layers.  In the  large  slitlike pores the order parameter in the
surface layer  $\Delta\rho_{1}$($\tau$) demonstrates a  universal
temperature dependence,  consistent with the exponent $\beta_1$ = 0.82
of the  ordinary transition (figure 11,a). In  the  second and
subsequent layers,  $\Delta\rho_{i}$($\tau$) shows a crossover from
bulk to surface  critical behaviour  with increasing temperature
(figure 9). The increase of the crossover temperature with the
distance to the surface agrees with the expectation, that intrusion
of the surface perturbation into   the    bulk   is governed   by
the bulk correlation   length \cite{Binderrev1}.  As a  result,
the effective  critical  exponent $\beta^{eff}$ varies smoothly
from the bulk value in the pore interior to  the value $\beta_1$
near the  surface  (figure 10).  A very  similar crossover
behaviour was observed  for the ordinary  transition in Ising magnets
\cite{Rovere,Selke,Pleimling2,Bengrine}.
\begin{figure}[ht]
\begin{center}
\includegraphics[width=10cm]{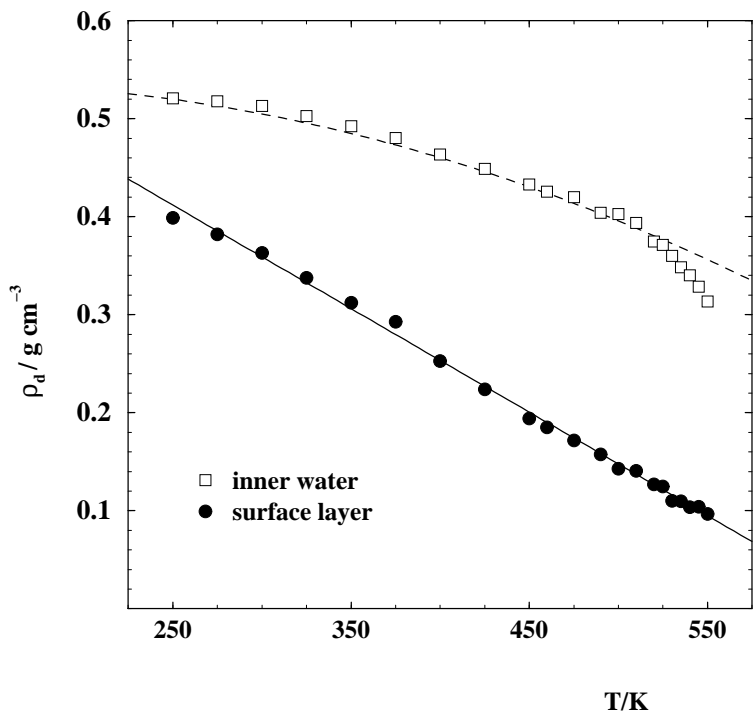}
\caption{The temperature dependence of the diameter $\rho_d$ for surface
layer   and  inner   water   in  a   slitlike   pore  with   H  =   30
$\mbox{\AA}$. The solid  line is a linear  fit of the diameter  in the surface
layer. The diameter  of bulk TIP4P water  \cite{BGO2004} is shown by
the dashed line.}
\end{center}
\end{figure}
  \par Figure 11 shows that in the three largest  slitelike  pores a  power
 law with  exponent
$\beta_1$ $\approx$  0.8 is valid  for the  surface layer in  an
extremely wide temperature  range (about  250 K).  This  means
that  the surface behaviour in  these pores is close  to that in
semiinfinite systems and the influence  of the opposite  wall is
negligible up  to a few  degrees from T$_C^{pore}$.  In smaller pores
the effective value of $\beta_1$ shows a  rapid increase, when approaching
T$_C^{pore}$. This  effect is more pronounced and
starts at lower temperatures in narrower pores. This  evidences
the influence of  the  opposite wall  on the  surface critical
behaviour.
\par   In  cylindrical  pores  the  temperature
evolution  of the  oder parameter  is in  general similar  to that
in slitlike pores.  Namely, the effective critical exponent
$\beta^{eff}$ increases   when  approaching the  surface  and
achieves values, which are essentially higher than the bulk  critical
exponent (figure 12, see also figure 5 in \cite{BGOPCCP}).
However, contrary  to the  planar surface,  the critical behaviour
of the order parameter in the surface layer even in large  pores
shows  a sharp  temperature crossover  from  the exponent
$\beta_1$ $\approx$ 0.8 to a much  higher value (figure 12). This
effect is so strong, that it can  be seen as a pronounced  shoulder
on the liquid branch and  diameter of  the coexistence curve  of
water in the pores with radius  R  =  25  $\mbox{\AA}$
(figure 2), R = 15  and  20 $\mbox{\AA}$ (figure 12 in
\cite{BGO2004}). This shoulder seems to shift to  higher
temperature  with increasing pore  radius.   The  order parameter
in  the second and subsequent layers do not  show a behaviour with
an exponent $\beta_1$ $\approx$ 0.8, rather a crossover directly from the
bulklike behaviour to  the behaviour with $\beta_1$
$\approx$ 1.8 occurs (figure 12).  This phenomenon is obviously
caused by the deviation of the surface from  planarity. The critical
behaviour of  the order parameter near  cylindrical surfaces is not
yet studied. Our results suggest  that a power  law critical
behaviour could also be valid  for such geometries. The obtained value of
$\beta_1$ $\approx$ 1.8 is comparable with the value 1.86 for the
corner magnetization of a cube, and to values observed  for edges
with an opening  angle  less  than  $\pi$/2 \cite{Pleimling}.
\par      The     local     order     parameter
$\Delta\rho$(z,$\tau$),  used  in  the  present  paper,  was
defined similarly as for the bulk fluid  (see above).  Like  in the
bulk  case, the order parameter is the  deviation of
the density  from the diameter. In the fluid near the surface
the diameter depends not only on  temperature, but also on the
distance from the surface. Such a definition of $\Delta\rho$(z,$\tau$)
provides a vanishing local  order
parameter at the critical point at any distance from the surface,
i.e.  the  profile of the order parameter is flat
($\Delta\rho(z,$ $\tau$ =  0) $\equiv$ 0), exactly as in the case of
the ordinary transition of Ising magnets \cite{Binderrev1}.  In
large pores the temperature dependence of the diameter of the coexistence curve in
the pore interior  is close to the  bulk behaviour up to the
temperature, where the  "inner" water  is influenced by  the
surface  (figure 14).  In the surface  layer the local  diameter
$\rho_d$ shows a perfect linear temperature dependence  in the
whole temperature range (figure 14). The critical  density in the surface
layer decreases with increasing pore size and achieves an
extremely low value (0.06 g cm$^{-3}$ in slitlike pore with H = 30
$\mbox{\AA}$).  This  results in  a  decrease of the average  pore
critical density with  increasing pores size (figure 5) and may
indicate a drying transition at the critical point of a
semi-infinite system.

\par  The   profiles  of  $\Delta\rho$(z,$\tau$) could   provide  additional
information concerning  the surface  critical behaviour of  fluids.  The
profile of magnetization was  studied in mean-field approximation both
for the ordinary  and extraordinary transitions \cite{Kumar}. In the case of a
ordinary  transition (h$_1$  = 0),  the local  magnetization  is symmetrical
in the two phases and therefore serves as order parameter. Near the
surface it obeys the  following  dependence  on the  distance  z to  the
surface \cite{Binderrev1,Diehlrev}:
\begin{eqnarray}
    \Delta\rho(z) = \Delta\rho_{bulk} \tanh(z/(2\xi_-) + z^*),
\end{eqnarray}
   where $\Delta\rho_{bulk}$,  the value of the  bulk order parameter,
and $\xi_-$, the bulk correlation length (T $<$ T$_C$), depend only on
temperature, while  $z^*$ is connected  with the density profile  close to
the  surface and is determined  by the surface-fluid  interaction (in an Ising
system it  is directly related  to the so-called  extrapolation length
$\lambda$). The distance z = 0 was assigned to 1.25 $\mbox{\AA}$
from the wall, which corresponds  to the boundary of the volume occupied by water
 (see Methods). In reference \cite{Kumar} another equation was proposed:
\begin{eqnarray}
\Delta\rho(z) =  \Delta\rho_{bulk}\frac{\sqrt{1 + (\lambda/\xi_-)^2} -
1  +   (\lambda/\xi_-)\tanh(z/2\xi_-)}   {(\lambda/\xi_-)  +
(\sqrt{1 + (\lambda/\xi_-)^2} -1)\tanh(z/2\xi_-)},
\end{eqnarray}
which reduces to equation (7) for small $\lambda/\xi_-$.
\par  Depletion of water density near the surface
could be also analyzed in the framework of normal transition,
assuming the preferential adsorption of voids. As the  normal
transition in  the limit of h$_1 \rightarrow
 \infty$ is equivalent to the extraordinary transition \cite{Diehl}, the
preferential adsorption  in mean-field approximation could be
described as \cite{Diehlrev,Kumar}:
\begin{eqnarray}
\rho(z) = a_1 - a_2 \coth(z/(2\xi_-) + z^*),~~when~T <
T_C
\end{eqnarray}
\begin{eqnarray}
\rho(z) = \rho_{bulk}  + a_1 /\sinh(z/\xi_+) +  z^*),~~ when~T >
T_C,
\end{eqnarray}

which provides an infinite adsorption at the  distance z =
-2$\xi_-$z$^*$  from the  surface. $\xi_+$ is  the bulk correlation length
at  T $>$ T$_C$. At subcritical temperatures we fitted equation (9) to the
density profiles of the coexisting phases and of the fluid order parameter
 $\Delta\rho$. In the latter case $\Delta\rho_{bulk}$ = (a$_1$ - a$_2$).

 \par A renormalization  group  analysis   of the  magnetization
profile  at  the ordinary transition  of an Ising  lattice shows,
that  very close  to the surface (z $\ll$ $\xi_-$) the order
parameter profiles obey a power law dependence $\Delta\rho$(z)$\sim$
z$^{(\beta_{1}-\beta)/\nu}$ and at z $\approx$    $\xi_-$
crosses    over   to an exponential   behaviour \cite{Binderrev1}.
The deviation of the scaling behaviour from the mean-field profile
(equations (7) and (8))  is noticeable in the  range 0 $<$  z $<$
2$\xi_-$ and achieves its maximum value  of  about 4$\%$  at
z  $\approx$ $\xi_-$/2  \cite{Gom}. The presence of pronounced density oscillations
of fluids near surfaces even at high temperatures makes
it reasonable  to neglect these small corrections  and to use in  our  study the
 mean-field equations (7-10) to
fit the order parameter and density profiles.
\begin{figure}[ht]
\begin{center}
\includegraphics[width=10cm]{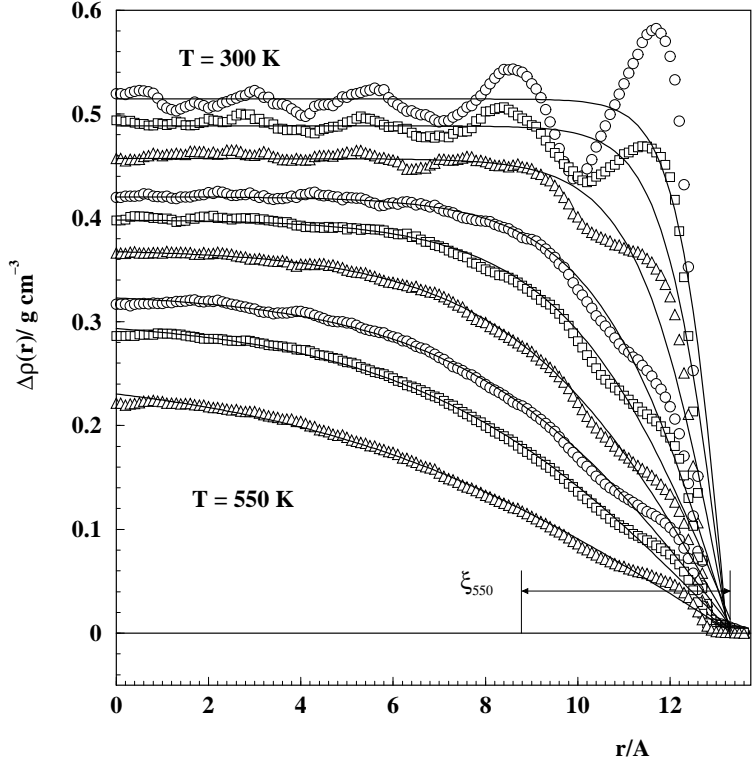}
\caption{Fitting  of the order  parameter profiles
$\Delta\rho$(r) in the slitlike  pore  with H  =  30  $\mbox{\AA}$  to
equations (7) and (8)  at  several temperatures: T  = 550, 535, 525,
500,  475, 450, 400, 350  and 300 K. The  value of  the
correlation  length, obtained  from the  fits, is shown for the
highest temperature.}
\end{center}
\end{figure}

\par  We  have fitted  the  profiles  of  $\Delta\rho$(z) as  well  as
$\rho_l$(z)  and  $\rho_v$(z) separately with equations (7-9)  and
the supercritical profiles of $\rho$(z) with equation (10),
assuming all parameters to  be freely variable. Equation (7)  and
equation (8)  provide equally good descriptions  of  the  order
parameter  profiles in  all studied slitlike pores  and in  the
whole temperature range  (see figure 15, where  fits using
equation (7) are shown). This evidences, that in our system the
extrapolation length $\lambda$ is negligibly small. Indeed, its
value practically does not depend on temperature and varies
between 0 and 0.5 $\mbox{\AA}$.  The order parameter
$\Delta\rho_{bulk}$, obtained from fits using equations (7) and
(8) are close to  the order parameter in bulk water at all
temperatures \cite{BGO2004}.  Small systematic  deviations of the
fitting curves  from the order  parameter in the pore interior,
which are noticeable  at   high  temperatures,  are  caused   by
deviations of $\Delta\rho$ in the pore interior  from the bulk
value discussed above and shown in figure 9.
\begin{figure}[ht]
\begin{center}
\includegraphics[width=10cm]{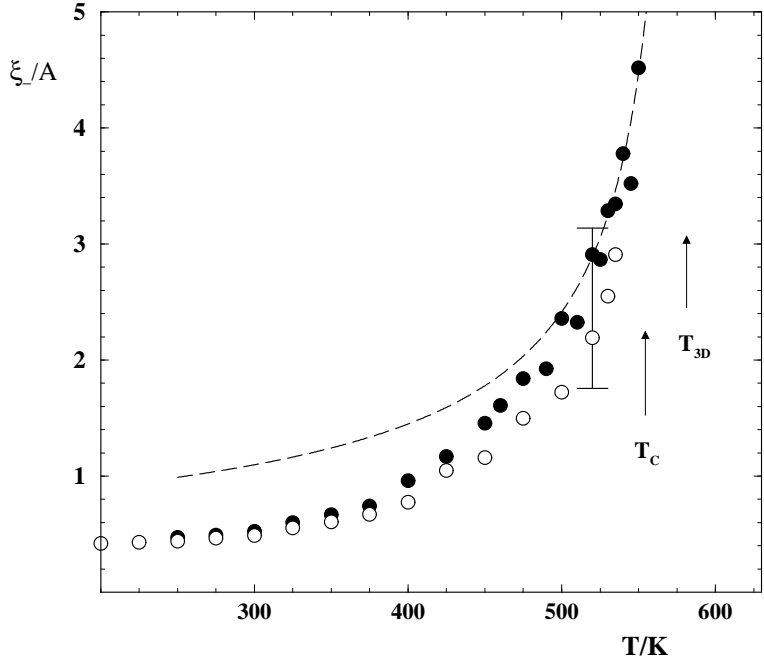}
\caption{Temperature  dependence of  the  effective   value  of
the correlation length along the coexistence curve obtained by
fitting equations (7) and (8) to the order parameter  profiles of the
slitlike pores with  H = 30 $\mbox{\AA}$ (solid  circles) and H =
24 $\mbox{\AA}$ (open circles). Simple  scaling   behaviour  of
the correlation  length   $\xi_-$  = $\xi_0$$\tau^{-0.63}$ with
the experimentally  obtained value $\xi_0$ = 0.694 $\mbox{\AA}$
\cite{Bonetti}, is shown by  dashed line. Vertical bar shows
increase  of $\xi_-$ at T = 520 K  with increasing pore size from
H = 18 $\mbox{\AA}$ to H = 50 $\mbox{\AA}$.}
\end{center}
\end{figure}\begin{figure}[ht]
\begin{center}
\includegraphics[width=10cm]{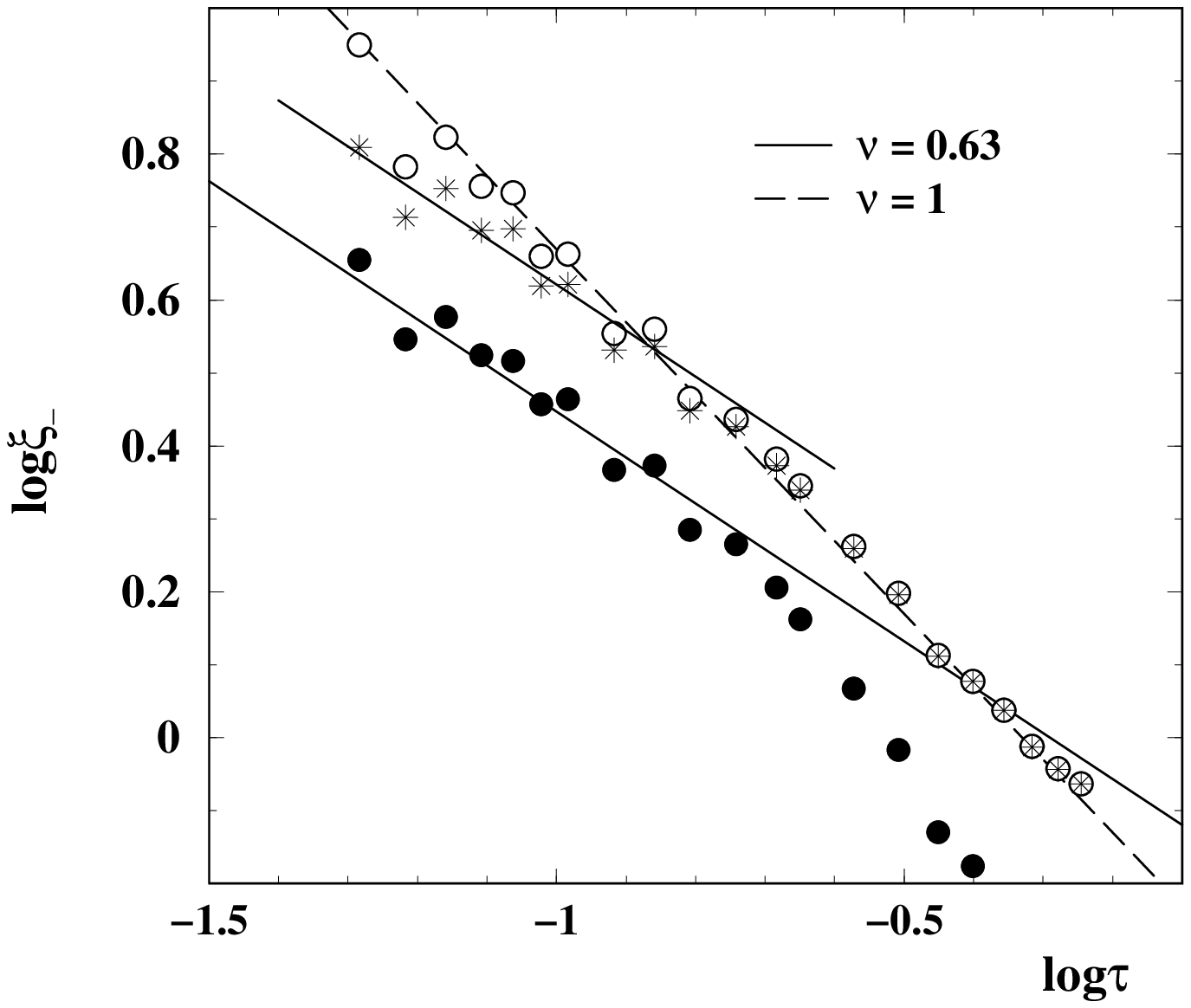}
\caption{Correlation lengths  along
the coexistence curve  of water in the  slitlike pore with  H =  30
$\mbox{\AA}$, obtained from: fitting equations (7) and (8) (solid circles) and
  equation (9) (open  circles) to the order parameter profiles; fitting equation (9) to the density profiles in the liquid phase (stars). }
\end{center}
\end{figure}

The temperature  dependence  of  $\xi_-$,
 obtained from  equations (7) and (8)
shows reasonable agreement with  the experiment (figure 16).
Moreover, at $\tau <$ 0.15 the correlation length shows the expected power
law behaviour $\xi_-(\tau$) $\sim$  $\tau^{-\nu}$ with  $\nu$ = $\nu_{3D}$ =
0.63 (figure 17). Note, that  the   values  of  $\xi_-$   were
obtained for a slitlike  (and not a semi-infinite) geometry and,
therefore, the correlation length $\xi_-$ increases with
increasing pore size.  For example, at T = 520 K $\xi_-$ =
1.76 $\mbox{\AA}$ in the pore with  H = 18  $\mbox{\AA}$, whereas in
the pore with H = 50 $\mbox{\AA}$ it achieves 3.14 $\mbox{\AA}$,
which seems to be close to convergence (see  vertical  bar in
figure 16) to the value in
a semi-infinite system (H $\rightarrow\infty$) $\xi_-$, i.e. to
the bulk correlation length. The progressive lowering of the effective $\xi_-$
with decreasing pore size reflects the shift of the phase
transition in the pore away from the bulk phase transition.
\par
Using equations (7) and (8) to describe the  order parameter
profiles in  cylindrical pores,  results  in increased effective
values of the correlation length.  As in the case of slitlike
pores, $\xi_-$ progressively lowers  with decreasing pore size.
However, the absolute value of $\xi_-$ considerably exceeds the
values of $\xi_-$ in slitlike pores of comparable sizes,
especially at high temperatures.
 Obviously, the order parameter  profiles in  cylindrical and
slitlike geometries  should be described by  different functions \cite
{Bindercyl} and the  critical behaviour of fluids  near cylindrical surfaces
deserves further  study.
  \par The fitting of equation (9) to the  order  parameter $\Delta\rho$(z)
is of lower quality in comparison with the fitting of
equations (7) and (8). The  obtained correlation lengths
follow an unexpected linear temperature dependence  in the double
logarithmic plot of figure 17.  So, the  order parameter
profiles $\Delta\rho$(z) show a behaviour, which is in accordance with
an ordinary transition.
 \par
The  density profiles in the liquid  phase $\rho_l$(z) in the
whole temperature range could be fitted with the same quality by any of the
equations (7-9).   The  values  $\xi_-$, obtained from the fits
using equations (7) and (8) are about 70 $\%$ of  the
values of $\xi_-$, obtained from
$\Delta\rho$(z).  The values of $\xi_-$, obtained from the fits
 using equation (9),  at low temperatures are practically
equal to the values $\xi_-$, obtained from  the corresponding  fits of
$\Delta\rho$(z), but  deviate noticeably  in  the
high temperature region, where the temperature dependence
becomes consistent with the expected scaling law (figure 17).
 Values of $\xi_-$, obtained  from the fit of the vapour density
profiles $\rho_v$(z) at high  temperatures, are several times  smaller
than the values $\xi_-$, obtained  from the fits of $\rho_l$(z) and
$\Delta\rho$(z).
\begin{figure}[ht]
\begin{center}
\includegraphics[width=10cm]{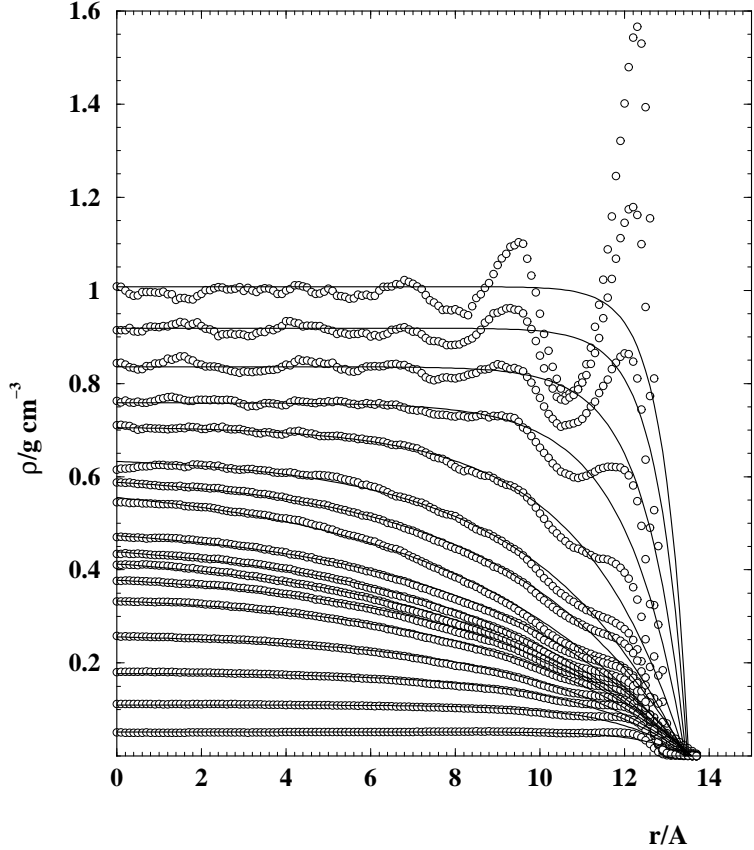}
\caption{Density profiles of supercritical water in the slitlike pore
with H =  30 $\mbox{\AA}$ at T = 580  K and at various  average
densities: from  0.047 g/cm$^3$  (bottom) to  0.945 g/cm$^3$
(top). The  fits using equation (10) are shown by lines.}
\end{center}
\end{figure}

\par The density  profiles of supercritical water (T = 580 K) at
several  average  densities  were  satisfactorily   fitted  by using
equation (10) (see figure 18), while the quality of the fits with
equation (7) was essentially worse.  The density  dependence of
the obtained correlation length  $\xi_+$ shows a
pronounced maximum, when the average density  is close to the
critical density of the bulk TIP4P  water $\rho$ $\approx$ 0.330 g
cm$^{-3}$ \cite {BGO2004} (figure 19).  A similar dependence was
observed experimentally for supercritical water \cite  {Japwater},
however, the absolute values of the correlation length, obtained
from simulations, are essentially smaller, as they are
obviously suppressed by the finite size of the simulated cells.
\par Summarizing our studies of
the order  parameter and  density profiles, we conclude: i) at  T
$<$ T$_C$ the profiles of  the order parameter $\Delta\rho$(z)
follows the behaviuor  of an ordinary  transition;  ii)  at T  $>$
T$_C$ the  density profiles  $\rho$(z)  follow the  behaviour  of
a normal transition  with preferential adsorption of voids.  The
studies of the density profiles of the coexisting  phases below
T$_C$ are not  conclusive.  Note also, that  in agreement  with
the simulation results for  LJ fluids \cite{Evans}, we never
observed a density minimum  in case of a preferential adsorption
of  voids, which could  be expected  for small  values of  the
surface field  \cite{Ritschel}.
\par  The  definition  of  the  local  order
parameter  as the  difference  of the local  densities  of the
coexisting phases,  introduced  in a  way  adopted  for  a bulk
fluid,  provides a vanishing  order parameter at T $\geq$
T$_C$ and therefore a flat profile  $\Delta\rho$(z) = 0  at T =
T$_C$.  This assumes  that the critical behaviour  of the  order
parameter $\Delta\rho$(z)  belongs to the   universality  class
of  ordinary   transition.    Indeed,  the temperature dependence
of the order  parameter in the surface  layer, its temperature
crossover in subsequent layers and the shape of the order
parameter profiles follow the behaviour of an ordinary transition.
\par
The densities  of the coexisting  phases $\rho_{l,v}$ in  general case
could be written as:
\begin{eqnarray} \rho_{l,v}(z,\tau) = \Delta\rho(z,\tau)+
\rho_d(z,\tau),
\end{eqnarray}

  where a symmetrical  contribution is presented by  the order parameter
  $\Delta\rho$(z,$\tau$), while the asymmetric term $\rho_d$(z,$\tau$) is
  the diameter, which contains regular as well as some singular terms.
  At  the critical  point  the order  parameter  vanishes and the density
  profile  reflects  a  preferential  adsorption, which  is  described
  solely by  the asymmetric  term $\rho_d$(z,$\tau$).  We  assume that
  equation (2) for  the critical behaviour in  the surface layer  in the case of
  a normal transition below T$_C$ describes the asymmetric contribution
  to the density profiles.  So,  the densities of the
  coexisting phases  in the surface layer (z  = 0) below  T$_C$
should behave in general as superposition of equation (1) and (2):
\begin{eqnarray} \rho_{l,v}(z=0,\tau) = (\rho_{1C} + A_1 \tau +
A_2\tau^2 + ...) + B_- \tau^{2-\alpha} \pm B_1 \tau^{\beta_1}
\end{eqnarray}
     In  the case  of  coexisting bulk liquid and  vapour phases  (z
 $\rightarrow  \infty$) the exponent  $\beta_1$ =  0.82 in equation (12)  should be
 replaced by $\beta$ = 0.326, $\rho_C$(z) by the bulk critical density
 $\rho_{3D}$ and the  exponent (2 - $\alpha$) by  the critical anomaly
 exponent of the diameter (1  - $\alpha$) \cite {Mermin}.  (Note, that the
 exponent $\alpha_{ord}$  for the ordinary transition  is $\alpha_{ord}$ =
 $\alpha$ -  1 \cite  {Diehlrev}, providing 1 - $\alpha_{ord}$ = 2 -
 $\alpha$). The validity of the critical anomaly of a diameter
 $\tau^{(1  - \alpha)}$ in fluids is questionable  due  to the so called pressure
 mixing  in  the scaling  fields,  which
 generates a term $\tau^{2\beta}$, that could dominate the $\tau^{(1 -
 \alpha)}$ term when $\tau \rightarrow 0$ \cite{Fisher2b}.

\begin{figure}[ht]
\begin{center}
\includegraphics[width=10cm]{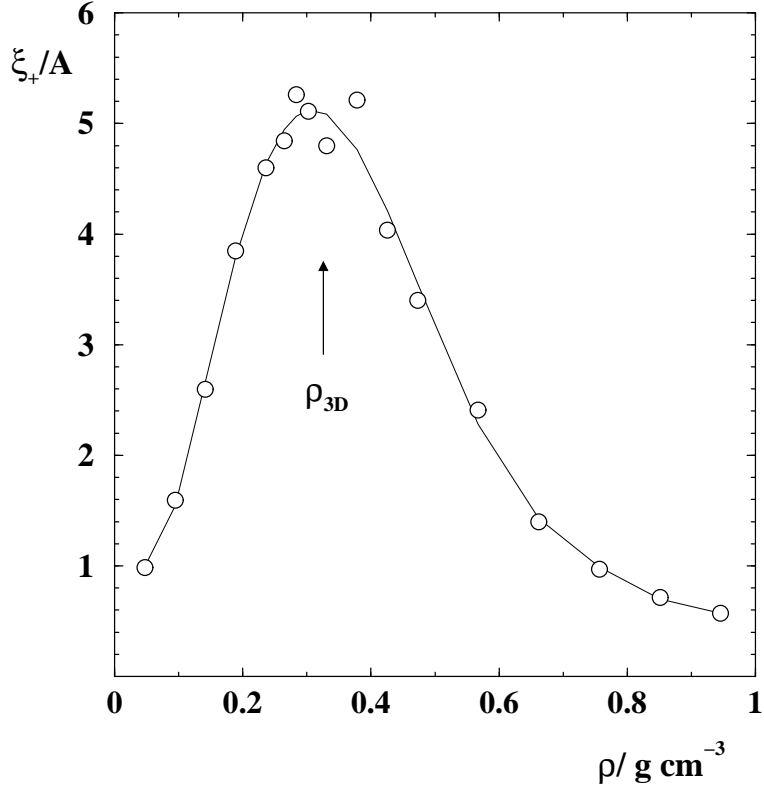}
\caption{Effective values of the correlation length along the supercritical
isotherm T  = 580  K,  obtained from  the  fit of  the
supercritical  densities profiles, shown  in figure 16,  to
equation (10) Critical  density of  the bulk TIP4P water
$\rho_{3D}$ = 0.330 g/cm$^3$ \cite{BGO2004} is  shown by arrow.
Line is a guide for eyes only.}
\end{center}
\end{figure}
\par To our knowledge, the temperature dependence of  the difference
 between the magnetizations of the coexisting phases
near the surface $\Delta$m = (m$_1$ - m$_2$)/2  is not studied for the
Ising model, when h$_1$ $\neq$ 0. At any non-zero value of a
surface field the $\Delta$m near the surface vanishes below T$_C$
at the wetting transition temperature \cite{NF82}. So, the
occurrence of the wetting transition in the Ising model with
short-range non-zero surface field h$_1$ and in some one-component
fluids or fluid mixtures prevents an analysis of the asymptotic
behaviour of the order parameter, defined as the difference between
magnetizations (densities, concentrations) of the coexisting
phases.  However, we may assume, that below the temperature of the
wetting transition the equation (12) is valid. This agrees with
the available experimental studies of the local order parameter in
binary mixtures \cite{Fenzl,Franck}. When the long-range
fluid-wall interaction makes a wetting (drying) transition
impossible at any T $<$ T$_C$ (the case studied in the present
paper), equation (12) should be valid also asymptotically  at
$\tau$ $\rightarrow$ 0.
 \par  Nevertheless, it is not clear, whether equation (12) is valid for any
 value of  h$_1$ or it is valid in our studies due the small value of  h$_1$.
 The  extremely  weak fluid-wall  interaction
 (about 10  $\%$ of fluid-fluid interaction)  provides practically the
 strongest  possible value  h$_1$ for  the preferential  adsorption of
 voids for existing solid substrates.  Assuming that the case  h$_1$ =  0 is
 signaled  by  the  flat  density  profile  at  the  critical  point
 \cite{Evans}, we can expect that equation (12) will be valid for the water-wall
interactions with  well-depths U$_0$ from -0.39 kcal/mol (present studies)
 to about -1.0  kcal/mol \cite{BGO2004}. Further  strengthening of the  fluid-wall
interaction
 causes a change of the sign  of the  surface field  h$_1$ in favour of
 molecules. However, the possibility to increase this field is limited
 by the formation of dead layers at the wall \cite {BGO2004}, which are identical
 in  both  coexisting  phases  and, therefore,  effectively  form a  new
 liquid-like wall. As  a  result the  fluid-wall interaction  comes
 close to the fluid-fluid interaction and is only slightly influenced by
 the interaction between molecules and solid wall.  Note also, that a
 possible  change  of h$_1$  with  temperature  \cite {Kisel}  can
 not be excluded and may complicate  the study  of  the critical  adsorption in  one-component fluids \cite  {Findenegg,Chanads}.
\par  So, the presented  study of
 water  near a hydrophobic  surface in  a wide  temperature range below
 T$_C$  and in the supercritical  region shows,  that the behaviour  of the
 order parameter $\Delta\rho$ is described by the laws of the ordinary
 transition,  while the  behaviour of the  asymmetric contribution  to the
 densities  of the liquid ($\rho_l$)  and vapour  ($\rho_v$)  phases is
 consistent  with the  laws  of the  normal  transition.  The  surface
 critical behaviour of a one-component fluid for various strengths of
 fluid-wall interaction  and in closer proximity to  the bulk critical
 temperature definitely needs further investigations.
\ack
This  work  was  supported  by Deutsche  Forschungsgemeinschaft.   The
authors  thank   K.Binder,  H.W.Diehl,  S.Dietrich   and  R.Evans  for
stimulating discussions and useful advices.

\end{document}